\documentclass[letterpaper]{article}
\usepackage{graphicx,subfigure}
\usepackage{cite}
\usepackage{ifpdf}
\usepackage{amsmath,amssymb,amsfonts}
\usepackage{textcomp}
\usepackage{algorithmic,algorithm}
\usepackage[margin=1in]{geometry}
\usepackage{multirow}
\usepackage{tabularx}

\title{Deep Gaussian Process-Based Bayesian Inference for Contaminant Source Localization}
\author{Young-Jin Park, Piyush M. Tagade, and Han-Lim Choi} 

\begin{document}
\maketitle
\begin{abstract}
	This paper proposes a Bayesian framework for localization of multiple sources in the event of accidental hazardous contaminant release. 
	The framework assimilates sensor measurements of the contaminant concentration with an integrated multizone computational fluid dynamics (multizone-CFD) based contaminant fate and transport model.
	To ensure online tractability, the framework uses deep Gaussian process (DGP) based emulator of the multizone-CFD model. 
	To effectively represent the transient response of the multizone-CFD model, the DGP emulator is reformulated using a matrix-variate Gaussian process prior.
	The resultant deep matrix-variate Gaussian process emulator (DMGPE) is used to define the likelihood of the Bayesian framework, while Markov Chain Monte Carlo approach is used to sample from the posterior distribution.
	The proposed method is evaluated for single and multiple contaminant sources localization tasks modeled by CONTAM simulator in a single-story building of 30 zones, demonstrating that proposed approach accurately perform inference on locations of contaminant sources.
	Moreover, the DMGP emulator outperforms both GP and DGP emulator with fewer number of hyperparameters.
\end{abstract}

\section{Introduction}
\label{sec:introduction}
Modern smart buildings are equipped with sensor networks and Internet-of-Things (IoT) technologies that monitor the daily indoor events, and store a significant amount of generated data~\cite{abraham2014cost, kim2014issaq, chen2014indoor, kumar2016indoor}.
In the state of the art smart buildings, however, the sensor data is primarily used to alert the occupants in the event of an accident. 
As demonstrated in this paper, the sensor data can play a more critical role in ensuring the occupant safety. 
For example in the event of an accident, the recorded data can be used to infer the causality and localize the accident source. 
Subsequently, this information can be used to make intelligent decisions like effectively controlling the heating, ventilation and air conditioning (HVAC) systems and planning the safest and fastest evacuation plan. 
Thus to effectively ensure the occupant safety, the smart building requires an accurate and fast inference algorithm. 
The main aim of this paper is to develop an online Bayesian inference framework~\cite{kennedy2001bayesian, tagade2009bayesian} for smart buildings. 
This Bayesian framework is developed for a specific event; an accidental (or deliberate) hazardous contaminant release in an indoor environment.

In the event of indoor hazardous contaminant release, sensor network monitors the concentration of pollutants like carbon monoxide (CO) gas that may negatively impact the occupant health \cite{spengler1983indoor, jones1999indoor, raub2000carbon}.
Due to diffusion, the contaminant disperses to other rooms in the building. 
This contaminant dispersion is modeled using a contaminant fate and transport model~\cite{chen2008sensor, chen2010comparison}.
As the contaminant spreads to different rooms, sensors in these rooms record the contaminant concentration.
These multiple sensor measurements are used for source localization.
One possible approach for source localization is to consider the room with the highest contaminant concentration as a target region. However, this approach fails to account for the concentration variation due to the spatial location of the source inside the room. 
Moreover, this approach fails to fully utilize the sensor data, and more critically, can not account for the uncertainty emanating from the sensor noise.

An alternative source localization approach is to assimilate the sensor data with the contaminant fate and transport model.
Traditionally the data assimilation approaches are implemented using deterministic algorithms based on optimization \cite{mahar1997optimal}, Kalman filtering \cite{federspiel1997estimating}, and inverse modeling \cite{liu2007inverse, zhang2007identification}.
However, these approaches do not provide full uncertainty analysis of contaminant source location and characteristics.
Primarily due to their ability to fully quantify the uncertainties, the Bayesian framework for contaminant source localization and characterization is gaining prominence in various studies\cite{keats2007bayesian, sreedharan2007bayesian, sohn2002rapidly, tagade2013gaussian}.
However, barring few notable exceptions of conjugacy like linear models with Gaussian priors, estimation of the Bayesian posterior distribution is analytically intractable.
Thus, the Bayesian framework is often implemented by approximating the posterior distribution using sampling methods like Markov Chain Monte Carlo (MCMC) \cite{hastings1970monte}.

Implementation of the MCMC based Bayesian inference framework for contaminant source localization and characterization is challenging due to: 1) for acceptable accuracy, MCMC requires $10^4-10^7$ samples \cite{keats2007bayesian, borysiewicz2012bayesian, wawrzynczak2013sequential, tagade2013gaussian}; 2) transient nature of the contaminant dispersion phenomena \cite{conti2010bayesian, tagade2013gaussian}.
Note that each MCMC sample requires simulating the contaminant fate and transport model, rendering the Bayesian framework computationally intractable for real-time contaminant source localization and characterization~\cite{keats2007bayesian}.
To resolve the first challenge, many state of the art approaches relies on simplification of the contaminant fate and transport model~\cite{sohn2002rapidly}.   
However, the computational cost of these simplified models is still prohibitive for online applications. 
Moreover, these models use model order reduction which results in a loss of spatial information. 

Alternatively, recent studies have explored various machine learning algorithms to develop computationally efficient emulators to infer the indoor events~\cite{manic2016intelligent}.
A similar approach is adopted in this paper, where the contaminant fate and transport model is replaced by a statistical emulator in the Bayesian framework.
Notwithstanding their relative merits, this paper focuses on Gaussian process based emulator \cite{conti2010bayesian} for the contaminant fate and transport model.
In earlier studies, Gaussian process emulator (GPE) has achieved state of the art results for contaminant source localization and characterization in a small single-story building \cite{tagade2013gaussian}. 
In this paper, the GPE approach is extended for contaminant source localization and characterization in a large building.
However, the Gaussian process (GP) requires embedding in an additional probabilistic structure (e.g. highly parameterized kernels) to learn accurate representation of a complex data, imposing higher computational cost~\cite{damianou2015deep, salimbeni2017doubly}.
Thus, GPE is no longer appropriate for the larger scale problems as the data complexity increases with the increasing size of the building and the number of rooms.

Recently developed deep variant of the GP, known as deep Gaussian process (DGP)
~\cite{bui2016deep, damianou2015deep, damianou2013deep, salimbeni2017doubly}, allows an accurate representation of the complex data at a reduced computational cost with fewer number of hyperparameters to learn.
However, usage of DGP emulator (DGPE) in the context of contaminant fate and transport model is not yet reported in the literature.
First key contribution of this paper is exploration of the DGP as a computationally efficient statistical emulator of the contaminant fate and transport model. 

Although the DGPE resolves first challenge in implementation of the MCMC based Bayesian inference algorithm, the existing DGP algorithm does not consider correlation in multi-output response (i.e. covariance between outputs), thus the DGP approach fails to efficiently represent the transient phenomenon. 
There were several researches to effectively deal with multi-output single-layer GP by using convolutional process \cite{boyle2005dependent, alvarez2009sparse}, considering the covariance matrix among multi-output \cite{bonilla2008multi, conti2010bayesian}, or sharing multiple sets of inducing variables \cite{nguyen2014collaborative}.
Among them, the matrix-variate approach is used in the literature to represent the transient phenomenon~\cite{gupta1999matrix, conti2010bayesian, tagade2016bayesian}.
The second key contribution of this paper is a generalization of the DGP formulation for matrix-variate Gaussian process (MGP), the resultant formulation is termed in this paper as a deep matrix-variate Gaussian process emulator (DMGPE).

Unlike single-layer GP, the DGP are no longer GP, thus the inference of marginal likelihood of the DGP is intractable. This causes challenges at estimating the objective function of DGP to maximize. To approximate the likelihood, Damianou et al. \cite{damianou2013deep} used variational approaches based on mean-field approximation, and optimized the evidence lower bound (ELBO) alternatively. Bui et al. \cite{bui2016deep} applied expectation propagation (EP) approach to improve the performance of DGP at regression tasks. Most recently, Salimbeni et al. \cite{salimbeni2017doubly} introduced doubly stochastic variational inference method that assumes neither independence  between GP layers nor Gaussianity for GP outputs. In this paper, we extended the work from \cite{salimbeni2017doubly} to the matrix-variate formulation without using any mean-field variational approximation. As such, the matrix-variate variational distribution can recover the true posterior in a more accurate way. Furthermore, we take the importance weighted auto-encoder (IWAE) bound \cite{burda2015importance}, a family of Monte Carlo objectives (MCO) \cite{mnih2016variational}, that is known to give a tighter lower bound compare to the traditional ELBO.

Note that the term deep matrix-variate Gaussian process is used earlier in~\cite{louizos2016structured}. 
However, the work presented in~~\cite{louizos2016structured} primarily focusses on neural networks with uncertain weights and biases.
A matrix-variate Gaussian distribution is used to specify prior on these uncertain weights and biases, while the variational inference is used to approximate the posterior by another matrix-variate Gaussian distribution. 
As the resultant network approximates the multi-output deep Gaussian process in the limit as the number of hidden nodes increase to infinity, this neural network is termed in~\cite{louizos2016structured} as a deep matrix-variate Gaussian process. 
Whereas in the present work, the matrix-variate Gaussian process is used to specify the prior on the transient model output. 
Subsequently, parameters of this matrix-variate Gaussian process are estimated using the training data.

Rest of the paper is organized as follows. 
Section~\ref{sec:bgd} provide the background of the key concepts used in this paper.
This section summarizes contaminant fate and transport model, the Bayesian framework for source localization and the DGP formulation. 
Section~\ref{sec:dmgpe} provide details of DMGP emulator (DMGPE) formulation and the optimal implementation using the approximate inference method.
Section \ref{sec:contam} describes the proposed Bayesian inference algorithm for source localization.
In section~\ref{sec:results}, an efficacy of the proposed framework is demonstrated for case studies involving contaminant source localization in a single-story building with 30 zones. 
Finally, the paper is summarized and concluded in section~\ref{sec:con}.

\section{Background} \label{sec:bgd}
\subsection{Contaminant Fate and Transport Model}
Multizone model is a widely used, simplified indoor contaminant fate and transport model~\cite{sohn2002rapidly, sreedharan2007bayesian, liu2009prompt}.
The multizone model approximates a building as a network of well-mixed zones with homogeneous air properties and contaminant concentration.  
Each room and corridor is represented as a zone, while door and window act as airflow paths linking adjacent zones. 
Using the mass conservation under the steady state condition inside the zone $j$, the flow rate between zone $i$ and $j$ is given by \cite{wang2007coupling}:
\begin{align} \sum_i C_i F_{ij} + \sum_j C_j F_{ji} + S_j = 0 \nonumber \\
F_{ij} = c_{ij} \mathrm{sign}(\Delta P_{ij}) \Delta P_{ij}^{n_{ij}}
\end{align}
where $F_{ij}$, $c_{ij}$, $\Delta P_{ij}$, $n_{ij}$, $C_i$, $C_j$, $S_j$ denotes flow rate, flow coefficient, flow exponent, pressure difference, contaminant concentration, and contaminant source in the zone $i, j$, respectively.

Although simple and computationally efficient, the multizone model does not provide an accurate airflow distribution. 
When the spatial contaminant distribution inside a zone is required, a computational fluid dynamics (CFD) is more appropriate for contaminant fate and transport modeling.
CFD numerically solves a set of partial differential equations for mass, momentum and energy conservation to obtain transient airflow rate among zones \cite{wang2007coupling}:
\begin{equation} \nabla(\rho V u)  - \Gamma_u \nabla^2u = S_u \label{CFD}\end{equation}
where $u, V, \rho, \Gamma_u, S_u$ is conservation variable, density, velocity, diffusion coefficient, and the source, respectively. 
Although accurate, CFD model is computationally very expensive.

Current state of the art contaminant fate and transport models integrate the multizone and CFD model to exploit both the advantages of efficient computation and accurate airflow modeling~\cite{tan2005application,wang2007coupling, kim2015development}.
Coupling between multizone and CFD model is achieved by establishing the convergence of total mass flow for the every flow path within a building. 
This paper utilizes the integrated multizone-CFD model developed by CONTAM~\cite{wang2007coupling, dols2015development, dols2015contam} for contaminant fate and transport simulation.

\begin{figure}[t!]
	\includegraphics[width=\columnwidth]{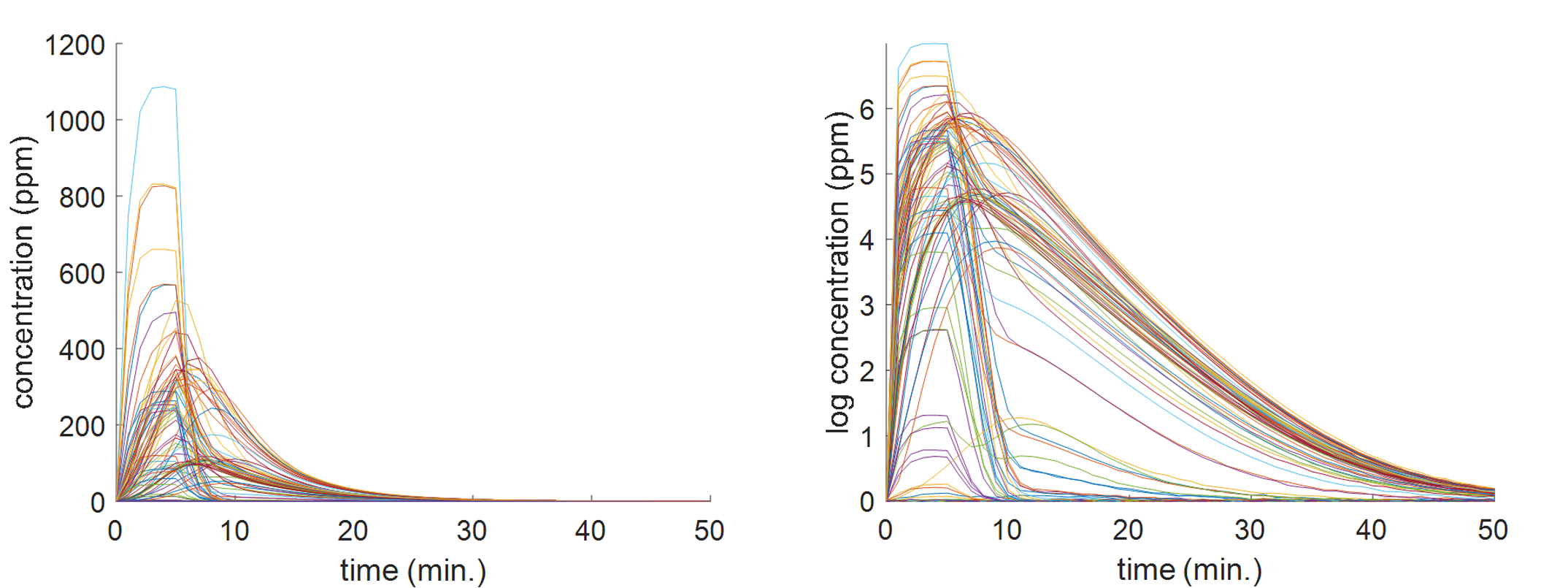}
	\caption{Unprocessed (left) and log transformed (right) contaminant concentration curves at different source locations.}
	\label{log}
\end{figure}

\subsection{Bayesian Framework for Source Localization}
Let the transient response of the multizone-CFD model at the time instances $\mathcal{T}=\{t_k; ~~~ k=1,\cdots,T\}$, and let outputs of multizone-CFD model at time points $\mathcal{T}$:
\begin{equation}\zeta_j(\mathbf{x}) = \{C_j(\mathcal{T};\gamma, \mathbf{x})\}\label{MCFD}\end{equation}
where $\gamma$ is a set of deterministic inputs (e.g. room, flow path, outdoor condition, etc.), $\mathbf{x}$ is a set of uncertain parameters (e.g. the number, zone and location of contaminant source), and $\zeta_j$ is a transient contaminant concentration in $j^{th}$ zone. 
Note that since $\gamma$ and $\mathcal{T}$ are assumed to be completely known, explicit dependence of $\zeta_j$ on $\gamma$ and $\mathcal{T}$ is dropped in the notation. 

Let $\mathbf{x}^*$ be the true but unknown value of uncertain parameters.
The measurement from the sensor in $j^{th}$ zone is given by
\begin{equation}\beta_j = \zeta_j(\mathbf{x}^*) + \boldsymbol{\epsilon}_j ,\end{equation}
where $\boldsymbol{\epsilon}_j$ is sensor noise in the $j^{th}$ zone. Sensor noise is often assumed to be zero-mean Gaussian distributed:
$p(\boldsymbol{\epsilon}_j) = \mathcal{N}(\boldsymbol{\epsilon}_j; \mathbf{0}, \sigma_j^2\mathbf{I})$,
where $\mathbf{I}_P$ denotes an identity matrix and $\mathcal{N}(\mathbf{x}; \mathbf{m}, \mathbf{\Sigma})$ represents the probability density of vector $\mathbf{x}$ under the multi-variate normal distribution with mean vector $\mathbf{m}$ and covariance matrix $\mathbf{\Sigma}$.
Rather than using unprocessed sensor data, the log-transformed value may replace $\beta_j$ for the purpose of reducing skewness. In the present work, $\mathbf{y}_j \equiv \log(1+\beta_{j})$ transformation of sensor measurement data has been adopted as shown in Fig. \ref{log}.

Define a set of $N_s$ transformed sensor measurements $\mathbf{Y} = \{\mathbf{y}_j ; j = 1, \cdots, N_s\}$, and $\mathcal{Z} = \{\zeta_{j}(\mathbf{x}^*); j = 1, \cdots, N_s \}$ be a corresponding multizone-CFD model output for \emph{true} parameters $\mathbf{x}^*$.
Using Bayes' theorem, the posterior probability distribution of $\mathbf{x}^*$ is given by
\begin{align}
p(\mathbf{x}^*|\mathbf{Y}) & \propto p(\mathbf{Y}|\mathbf{x}^*)p(\mathbf{x}^*) \nonumber \\
~ & = p(\mathbf{Y}|\mathcal{Z} )p(\mathbf{x}^*).
\label{bayes}\end{align}
where $p(\mathbf{x}^*)$ is the prior probability distribution and $p(\mathbf{Y} |\mathbf{x}^*)$ is the likelihood. 
Assuming independent and identically distributed measurements, the likelihood is given by 
\begin{equation}
p(\mathbf{Y}|\mathcal{Z} ) = \prod^{N_s}_{j=1} p\left(\mathbf{y}_j | \zeta_j\left( \mathbf{x}^*\right) \right).
\end{equation}
The implementation of the Bayesian framework is completed by specifying $p\left(\mathbf{y}_j | \zeta_j\left( \mathbf{x}^*\right) \right)$ and sampling from the posterior $p(\mathbf{x}^*|\mathbf{Y})$; these steps are discussed in section~\ref{sec:dmgpe} and \ref{sec:contam}.

\section{Deep Matrix-Variate Gaussian Process Emulator} \label{sec:dmgpe} 
\subsection{Deep Gaussian Process}
Gaussian process is a flexible and powerful nonparametric Bayesian model for supervised learning \cite{rasmussen2006gaussian}. 
GP estimates the mean and covariance of the output at a given arbitrary input by projecting the correlation among the input space onto the output space. The deep Gaussian process is a statistical multi-layer hierarchical model of GP \cite{damianou2013deep}, which is designed for representation of highly complex data relationships. 
DGP has attracted wide attention owing to its superior performance for regression compared to other machine learning methods like single-layer GP and deep neural network \cite{damianou2013deep, bui2016deep, lawrence2007hierarchical}.

\begin{figure}[t!]
	\centering
	\includegraphics[width=0.5\columnwidth]{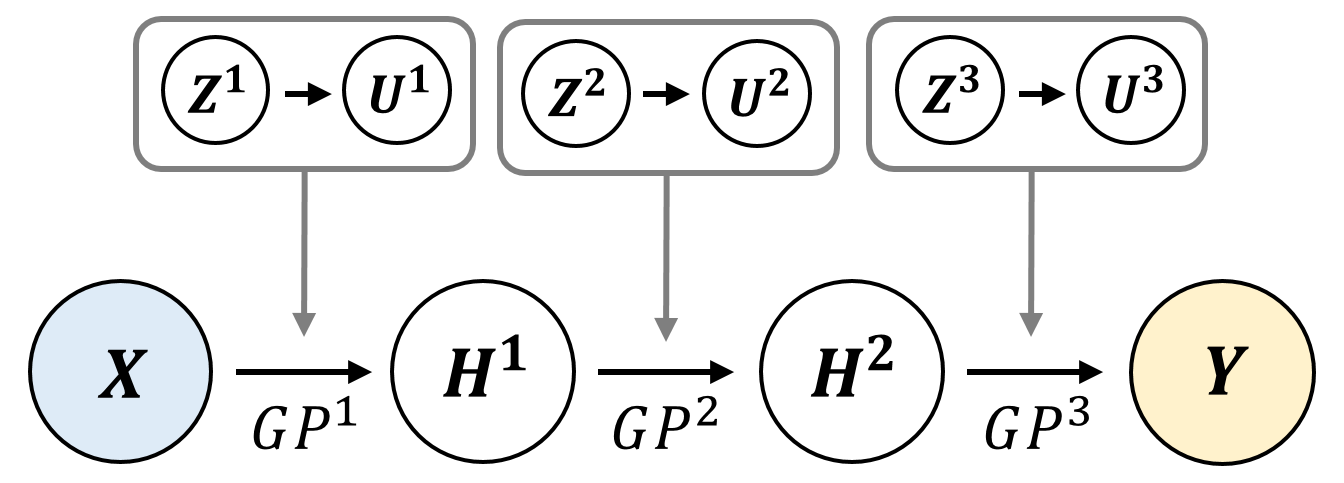}
	\caption{A graphical model of sparse DGP (DMGP).}
	\label{DGPgraphical}
\end{figure}

A graphical model of the DGP is shown in Fig. \ref{DGPgraphical}.
Consider a set of $N$ training data composed of $Q$-dimensional input $X = H^0 \in \mathcal{R}^{N \times Q}$, and $P$-dimensional output $Y = H^L \in \mathcal{R}^{N \times P}$. 
In general, DGP consisted of $L$-layer GPs is represented in a probabilistic way as follows:
\begin{align} \label{deep1}
p(\mathbf{F}^{l}) &= \prod_{d=1}^{D_l}\mathcal{N}(\mathbf{f}_d^{l}|m_d^l(\mathbf{h}^{l-1}), \mathbf{K}_{H^{l-1}H^{l-1}}) \nonumber \\
p(\mathbf{H}^{l}|\mathbf{F}^{l};\mathbf{H}^{l-1}) &= \prod_{n=1}^{N}\prod_{d=1}^{D_l}\mathcal{N}(h_{n,d}^{l}; f_{n,d}^{l},{w_d^l})
\end{align}
where $m^l(\cdot)$ is mean function, $\mathbf{F}^l$ is GP output matrix, $\mathbf{H}^l$ is observation matrix, $\mathbf{K}$ is GP kernel matrix, and $w^l$ is observation noise variance for $l^{th}$ hidden GP layer. Subscript $n$ denotes $n^{th}$ data set, and subscript $d$ denotes $d^{th}$ dimension of the variable.
Scalar, vector, and matrix values are, represented using lowercase italic, lowercase bold italic, and capital bold italic, respectively.
For instance, $\mathbf{F} = \{\mathbf{f}_n\}_{n=1}^N$, $\mathbf{f}_n = \{f_{n,d}\}_{d=1}^D$, and $\mathbf{f}_d = \{f_{n,d}\}_{n=1}^N$. Automatic relevance determination (ARD) squared exponential (SE) kernel is commonly used for the covariance matrix between $\mathbf{x}$ and $\mathbf{x}'$:
\begin{equation}\mathbf{K}_{\mathbf{x}\mathbf{x}'} = \sigma_f^2\exp{(-\frac{1}{2}\sum_d(\frac{\mathbf{x}_d-\mathbf{x}_d'}{\lambda_d})^2)}\end{equation}

To overcome extensive $\mathcal{O}(LN^3)$ cost of DGP, sparse approximation method is widely used \cite{snelson2006sparse, titsias2009variational, damianou2013deep, bui2016deep}. The method introduces $M$ inducing inputs ($\mathbf{Z}^l$) and corresponding outputs ($\mathbf{U}^l$) such that the computational complexity reduces to $\mathcal{O}(LNM^2)$. Assuming inducing variables are sufficient statistic for each GP layer, the resulting sparse model is given by
\begin{align} \label{deep2}
p(\mathbf{U}^{l}) &= \prod_{d=1}^{D_l}\mathcal{N}(\mathbf{u}_d^{l}|m_d^l(\mathbf{z}^l), \mathbf{K}_{Z^lZ^l}) \nonumber \\
p(\mathbf{F}^{l}|\mathbf{U}^{l};\mathbf{H}^{l-1}) &= \prod_{n=1}^{N}\prod_{d=1}^{D_l}\mathcal{N}(f_{n,d}^{l}; c^l_{n,d}, r_{n}^l) \nonumber \\
p(\mathbf{H}^{l}|\mathbf{F}^{l}) &= \prod_{n=1}^{N}\prod_{d=1}^{D_l}\mathcal{N}(h_{n,d}^{l}; f_{n,d}^{l}, {w_d^l}) 
\end{align}
where $\mathbf{c}_n^l = m^l(\mathbf{h}_n^{l-1}) + \mathbf{k}_{h_n^{l\!-\!1}Z^l}\mathbf{K}_{Z^lZ^l}^{-1}(\mathbf{U}^l - m^l(\mathbf{Z}^l))$ and
$r_{n}^l = k_{h_n^{l\!-\!1}h_n^{l\!-\!1}} - \mathbf{k}_{h_n^{l\!-\!1}Z^l}\mathbf{K}_{Z^lZ^l}^{-1}\mathbf{k}_{Z^lh_n^{l\!-\!1}}$.

\subsection{Deep Matrix-Variate Gaussian Process} \label{subsec:dmgpe_formul} 
Existing DGP model assumes probabilistic independence among output values in each dimension as shown in \eqref{deep1} and \eqref{deep2}, however this assumption is not appropriate for Bayesian emulation of transient response. 
Since each output of the transient response represents the value at a certain time instance, it is essential to consider the covariance in the temporal dimension.
Hence, this paper reformulates the DGP using the matrix-variate architecture~\cite{gupta1999matrix}.
For better clarity, this additional covariance is termed in this paper as output covariance matrix. 

Deep matrix-variate Gaussian process is represented as follows \footnote{see Appendix A for details}:
\begin{align} 
p(\mathbf{U}^{l}) &= \mathcal{MN}(\mathbf{U}^{l}; m^l(\mathbf{Z}^l), \mathbf{K}_{Z^lZ^l}, \mathbf{\Sigma}^l) \nonumber \\
p(\mathbf{F}^{l}|\mathbf{U}^{l};\mathbf{H}^{l-1}) &= \prod_{n=1}^{N}\mathcal{N}(\mathbf{f}^l_n; \mathbf{c}_n^l, r_n^l\mathbf{\Sigma}^l) \nonumber \\
p(\mathbf{H}^{l}|\mathbf{F}^{l}) &= \prod_{n=1}^{N}\mathcal{N}(\mathbf{h}^l_n; \mathbf{f}_n^l, \mathbf{W}^l) \label{DMGPform}
\end{align}
where $\mathbf{\Sigma}^l$ denotes output covariance matrix of $i^{th}$ hidden matrix-variate GP layer. $\mathcal{MN}(\mathbf{X}; \mathbf{M}, \mathbf{K}, \mathbf{\Sigma}) = \mathcal{N}(\mbox{vec}(\mathbf{X}); \mathbf{M}, \mathbf{\Sigma} \otimes \mathbf{K})$ is the probability density of matrix $\mathbf{X}$ under the matrix normal distribution with mean matrix $\mathbf{M}$ and two covariance matrices $\mathbf{K}$ and $\mathbf{\Sigma}$, where $\mbox{vec}(\mathbf{X})$ denotes the vectorization of $\mathbf{X}$ and $\otimes$ denotes the Kronecker product.
$\mathbf{W}^l$ is an observation noise covariance that is often assumed to be diagonal matrix. Notice that, DMGP model is equivalent to DGP model for the case when $\mathbf{\Sigma}^l$s are identity matrices, and single-layer DGP model is equivalent to GP model. That is to say, DMGP model is a generalized model encompassing DGP and GP, thus support the richer expressiveness.

To represent the output covariance matrix, kernelized \footnote{The covariance can be expressed by one-dimensional SE kernel: $\mathbf{\Sigma}_{ij} = \sigma_t \exp{(-\frac{1}{2}(\frac{t_i-t_j}{\lambda_t})^2)}$ where $\sigma_t$ and $\lambda_t$ are kernel hyperparameters to learn.} or free form convariance can be used.
This paper uses free form covariance approach as it can cover every possible symmetric positive definite matrix, thus providing higher degree of expressiveness as compared to the SE kernel;
SE kernel assumes constant time length scale along the time which is not the case in the contam problem.
Although free form covariance is difficult to be generalized for data with different time instances, we assume the time instances $\mathcal{T}$ for the emulator is fixed and try to learn the multi-output covariance that most represents the correlation for given $\mathcal{T}$.
Note, however, that the DMGP can be easily extended to kernelized covariance without any model modification.
By using Cholesky decomposition, $\mathbf{\Sigma}$ is parameterized as
\begin{equation}\mathbf{\Sigma} = \mathbf{L}_{\Sigma}\mathbf{L}_{\Sigma}^T \label{par}, \end{equation}
where hyperparameter $\mathbf{L}_{\Sigma}$ is a lower triangular matrix with positive diagonals.

\subsection{Approximate Bayesian Inference of DMGP}
Consider the log marginal likelihood of the model that is given by:
\begin{equation}
\log p(\mathbf{Y}) =  \log \int p(\mathbf{Y}, \mathbf{F}^L,\mathbf{U}^L, \{\mathbf{H}^l,\mathbf{U}^l\}_{l=1}^{L-1})\end{equation}
For the sake of notational simplicity, let $\bar{\mathbf{F}}^l \equiv \mathbf{H}^l$ for $l = 1, \cdots, L-1$ and $\bar{\mathbf{F}}^L \equiv \mathbf{F}^L$.
Using the variational inference method, we optimize the ELBO instead of analytically intractable log marginal likelihood of the model that is given by:
\begin{align}
\log p(\mathbf{Y}) &\geq \mathbb{E}_{q(\{\bar{\mathbf{F}}^l ,\mathbf{U}^l\}_{l=1}^L)}\left[\log \frac{p(\mathbf{Y},\{\bar{\mathbf{F}}^l,\mathbf{U}^l\}_{l=1}^L)}{q(\{\bar{\mathbf{F}}^l,\mathbf{U}^l\}_{l=1}^L)}\right] \nonumber\\
&\equiv \mathcal{L}(\mathbf{Y}), 
\end{align}
As we discussed in Section~\ref{subsec:dmgpe_formul}, the joint probability for $\mathbf{Y},\{\bar{\mathbf{F}}^l,\mathbf{U}^l\}_{l=1}^L$ is factorized by:
\begin{equation}
p(\mathbf{Y},\{\bar{\mathbf{F}}^l,\mathbf{U}^l\}_{l=1}^{L}) = p(\mathbf{Y}|\bar{\mathbf{F}}^L) \prod_{l=1}^{L}p(\bar{\mathbf{F}}^l|\mathbf{U}^l;\bar{\mathbf{F}}^{l-1})p(\mathbf{U}^l) \label{p_dist}
\end{equation}

In a similar way to \cite{hensman2013gaussian, salimbeni2017doubly}'s work, we choose variational distribution as:
\begin{equation}
q(\{\bar{\mathbf{F}}^l,\mathbf{U}^l\}_{l=1}^L) = \prod_{l=1}^{L}p(\bar{\mathbf{F}}^l|\mathbf{U}^l;\bar{\mathbf{F}}^{l-1})q(\mathbf{U}^l) \label{var_dist}
\end{equation}
where $\{q(\mathbf{U}^l) = \mathcal{MN}(\mathbf{U}^l; \mathbf{A}^l, \mathbf{S}^l, \mathbf{\Sigma}^l)\}_{l=1}^L$ are parameterized by variational parameters $\{\mathbf{A}^l, \mathbf{S}^l\}_{l=1}^L$ and output covariances $\{\mathbf{\Sigma}^l\}_{l=1}^L$.
Note that the variational distribution suggested does not impose independence between output dimension while existed approaches used mean-field variational approximation. Furthermore, by introducing output covariances, the number of hyperparameters is changed from $DM^2$ to $M^2 + D^2$ compare to existed DGPs.
Together with the contaminant problem, when $M^2$ is larger than $D^2$, DMGPs have approximately $D$ times smaller number of hyperparameters that enables faster model training.

Since terms in the variational posterior are matrix-variate normal, thus we can analytically marginalize \eqref{var_dist} over $\mathbf{U}^l$ for $n^{th}$ data:
\begin{align}
& q(\{\bar{\mathbf{f}}_n^l\}_{l=1}^L) = \int \prod_{l=1}^{L}p(\bar{\mathbf{f}}_n^l|\mathbf{U}^l;\bar{\mathbf{f}}_n^{l-1})q(\mathbf{U}^l) \nonumber\\
& = \mathcal{N}(\bar{\mathbf{f}}_n^L|\tilde{\mathbf{c}}_n^L, \tilde{r}_n^L\mathbf{\Sigma}^L) \prod_{l=1}^{L-1} \mathcal{N}(\bar{\mathbf{f}}_n^l|\tilde{\mathbf{c}}_n^l, \tilde{r}_n^l\mathbf{\Sigma}^l + \mathbf{W}^l). \label{var_marginal}
\end{align}
where $ \tilde{\mathbf{c}}_n^l = m^l(\bar{\mathbf{f}}_n^{l-1}) + \mathbf{k}_{\bar{f}_n^{l\!-\!1}Z^l}\mathbf{K}_{Z^lZ^l}^{-1}(\mathbf{A}^l - m^l(\mathbf{z}^l))$ and
$\tilde{r}_n^l = k_{\bar{f}_n^{l\!-\!1}\bar{f}_n^{l\!-\!1}} - \mathbf{k}_{\bar{f}_n^{l\!-\!1}Z^l}\mathbf{K}_{Z^lZ^l}^{-1}(\mathbf{K}_{Z^lZ^l}-\mathbf{S}^l)\mathbf{K}_{Z^lZ^l}^{-1}\mathbf{k}_{Z^l\bar{f}_n^{l\!-\!1}}$.

Following the ELBO derivation in \cite{salimbeni2017doubly}, the ELBO of DGMP is given by:
\begin{equation}
\mathcal{L}(\mathbf{Y}) = \sum_{n=1}^N \mathbb{E}_{q(\mathbf{f}_n^L)}\left[\log p(\mathbf{y}_n|\mathbf{f}_n^L)\right] - \sum_{l=1}^L KL(q(\mathbf{U}^l)||p(\mathbf{U}^l)) \label{elbo}.
\end{equation}

However, it is known that the traditional ELBO update heavily penalizes variational posterior samples, which fails to express the true posterior, that decreases the flexibility during a training procedure \cite{burda2015importance}. Instead, this paper utilized the MCO that is known to solve the above issue and known to give a tighter lower bound as well:
\begin{align}
\mathcal{L}^K(\mathbf{Y}) &= \sum_{n=1}^N \mathbb{E}_{q(\mathbf{f}_{n,k}^L)}\left[\log \frac{1}{K} \sum_{k=1}^K p(\mathbf{y}_n|\mathbf{f}_{n,k}^L)\right] \nonumber \\
& ~~~~ - \sum_{l=1}^L KL(q(\mathbf{U}^l)||p(\mathbf{U}^l)) \label{MCO1}
\end{align}
See Appendix B and C for detailed derivation of \eqref{var_marginal}, \eqref{elbo}, and \eqref{MCO1}.

In many cases, expectation term is difficult to compute analytically, thus it is often computed by Monte Carlo sampling technique \cite{gal2015latent, bonilla2016generic, salimbeni2017doubly}.
From the equation \eqref{var_marginal}, we can notice that $\bar{\mathbf{f}}_n^l$ depends only on the previous layer $\bar{\mathbf{f}}_n^{l-1}$, therefore:
\begin{equation}
q(\mathbf{f}_n^L) = \int \prod_{l=1}^{L} q(\bar{\mathbf{f}}_n^l; \bar{\mathbf{f}}_n^{l-1}) d\bar{\mathbf{f}}_n^{l-1}. \label{qfL}
\end{equation}
Thereby, we can easily sample $\mathbf{f}_n^L$ by recursively drawing samples from $q(\bar{\mathbf{f}}_n^l;\bar{\mathbf{f}}_n^{l-1})$ for $l = 1, \cdots, L$.
Using the fact that $q(\mathbf{f}_n^l)$ follows multi-variate Gaussian distribution, reparameterization \cite{kingma2013auto} technique can be applied:
\begin{align}
\mathcal{L}^K(\mathbf{Y}) &= \sum_{n=1}^N \mathbb{E}_{\boldsymbol{\epsilon}_k \sim \mathcal{N}(\mathbf{0},\mathbf{I})}\left[\log \frac{1}{K} \sum_{k=1}^K p(\mathbf{y}_n|\mathbf{f}_n^L(\boldsymbol{\epsilon}_k))\right] \nonumber \\
& ~~~~ - \sum_{l=1}^L KL(q(\mathbf{U}^l)||p(\mathbf{U}^l)) \label{MCO}
\end{align}
\begin{align}
\nabla_{\theta} \mathcal{L}^K(\mathbf{Y}) &= \sum_{n=1}^N \mathbb{E}_{\boldsymbol{\epsilon}_k \sim \mathcal{N}(\mathbf{0},\mathbf{I})}\left[\sum_{k=1}^K \tilde{\omega}_{n,k} \nabla_{\theta} \log p(\mathbf{y}_n|\mathbf{f}_n^L(\boldsymbol{\epsilon}_k))\right] \nonumber \\
& ~~~~ - \sum_{l=1}^L \nabla_{\theta} KL(q(\mathbf{U}^l)||p(\mathbf{U}^l)) . \label{MCOgrad}
\end{align}
where $\theta$ is parameters to optimize and $\tilde{\omega}_{n,k} = p(\mathbf{y}_n|\mathbf{f}_n^L(\boldsymbol{\epsilon}_k)) / \sum_{k=1}^{K} p(\mathbf{y}_n|\mathbf{f}_n^L(\boldsymbol{\epsilon}_k))$ are normalized importance weights.

\subsection{DMGP Emulator Implementation}
A primary purpose of the DMGPE is to find the approximating function that maps a given set of design points into transient data generated by the multizone-CFD simulator with reduced computational cost. 
Consider we perform $N$ simulation to generate training data. Define a set of design points as training inputs $\mathbf{X} = [\mathbf{x}_1, \cdots, \mathbf{x}_N]^T \in \mathcal{R}^{N \times Q}$ where $\mathbf{x}_n$ is a column vector representing locations of contaminant sources released at $n^{th}$ design point. Similarly, training output is defined as $\mathbf{Y}_j = [\mathbf{y}_{j,1}, \cdots, \mathbf{y}_{j,N}]^T \in \mathcal{R}^{N \times P}$ where $\mathbf{y}_{j,n}$ is a column vector of a sensor data at $p$ time points $\{t_1, \cdots, t_P\}$ in $j^{th}$ zone. 
Corresponding DMGPE $T_{j}(\cdot)$ estimates the sensor measurements in the $j^{th}$ zone with the multi-variate Gaussian. To simplify further notation, we omit subscript $j$ from here.

The performance of DMGPE relies on DMGP hyperparameters including kernel parameters, output covariances, inducing inputs, and variational parameters. 
Optimal DMPGE hyperparameters are obtained by maximizing the MCO of the model as follows:
\begin{equation}\theta_{opt} = \underset{\theta}{\arg\!\max}\mathcal{L}^K(\mathbf{Y}). \end{equation}
Implementation are provided in Algorithm \ref{emulator}. 
Note that derivatives of marginal likelihood for each DMGP hyperparameter are analytically computable, and the adaptive moment estimation (Adam) algorithm \cite{kingma2014adam}, one of the most widely used gradient-based optimization method, has been used in the present research work.
We implemented DMGPE with Tensorflow~\cite{abadi2016tensorflow}.

\begin{algorithm}
	\caption{Optimal DMGPE Implementation }
	\label{emulator}
	\begin{algorithmic}[1]
		\STATE Select $N$ design points for the experiment.
		\STATE Run $N$ simulations and generate training set ($\mathbf{X}$,$\mathbf{Y}$).
		\WHILE { $Iteration \le Maxiter$}
		\FOR {$k=1$ to $K$} 
		\FOR {$l = 1$ to $L$}
		\STATE Draw a sample from $q(\bar{\mathbf{f}}^l;\bar{\mathbf{f}}^{l-1}$)\footnotemark.
		\ENDFOR
		\STATE Store $\mathbf{f}^L$.
		\ENDFOR
		\STATE Compute the MCO $\mathcal{L}^K(Y)$ and the gradient $\nabla \mathcal{L}^K(Y)$ using (\ref{MCO}-\ref{MCOgrad}).
		\STATE Update DMGPE hyperparameters using Adam.
		\ENDWHILE
	\end{algorithmic} 
\end{algorithm}
\footnotetext{We take $K$ samples at once by batch processing using GPU.}

\begin{figure}[t!]
	\includegraphics[width=\columnwidth]{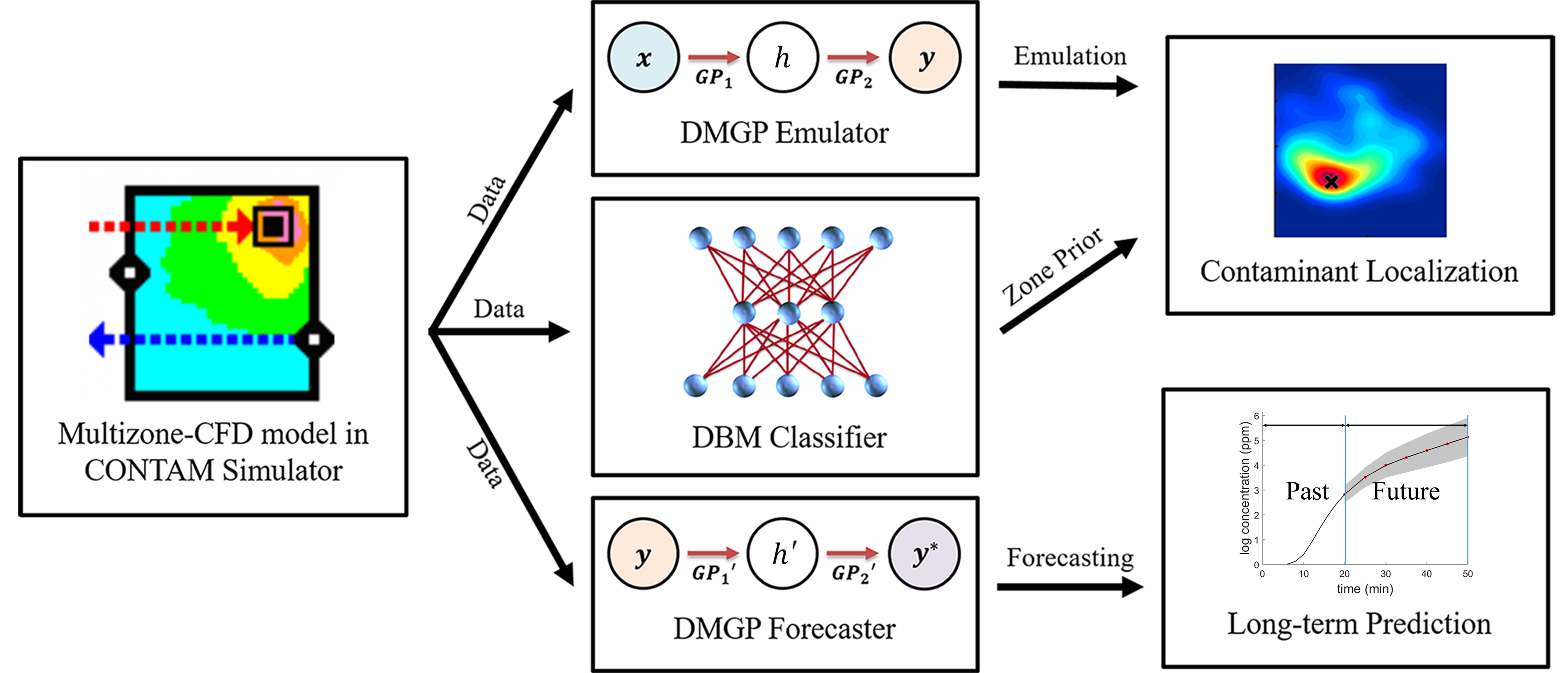}
	\caption{Flow diagram of solving contaminant problem using deep learning techniques.}
	\label{architecture}
\end{figure}

\section{Rapid and Accurate Source Localization} \label{sec:contam}
\subsection{Bayesian Source Localization using DMGPEs}
Consider an indoor building environment of $N_z$ zones equipped with $N_s$ sensors in $Z_s = \{z_1, \cdots, z_{N_s}\}^{th}$ zones. 
For effective emulators, DMGPEs $~T_{z,n,s}$ are built at each possible combination of $(z\!\in\!N_z, n\!\in\!N_c, s\!\in\!Z_s)$ where $z$ is the zone of active sources, $n$ is the number of active sources, $s$ is sensor location, and $N_c$ is possible numbers of active sources. In this study, $N_c$ are assumed to be 3. Active sources are assumed to be in the same zone since CONTAM simulator does not provide CFD model for multiple zones.

In this study, we focus on source localization problem, so the design point $\mathbf{x}^*$ is equivalent to the location of contaminant sources.
For a given design point $\mathbf{x}^*$, the DMGPE can sample $\mathbf{y}^*$ from $q(\mathbf{y}^*;\mathbf{x}^*)=p(\mathbf{y}^*|{\mathbf{f}^{L*}})q({\mathbf{f}^{L*}};\mathbf{x}^*)$ using \eqref{var_marginal}, \eqref{qfL} and \eqref{DMGPform}.
Then, transient contaminant concentration is approximated by multi-variate Gaussian distribution with sample mean $\boldsymbol{\mu}_s$ and sample covariance $\boldsymbol{\Omega}_s$:
\begin{equation}
T_{s}(\mathbf{x}^*) \sim \mathcal{N}(\boldsymbol{\mu}_s(\mathbf{x}^*),\boldsymbol{\Omega}_s(\mathbf{x}^*)).
\end{equation}
Consequently, the likelihood of observed data from the sensor network is given by
\begin{equation}
p(\mathbf{Y}|\mathbf{x}^*) \approx \prod_{s\in Z_s} p(\mathbf{y}_s|T_s(\mathbf{x}^*)) = \prod_{s\in Z_s} \mathcal{N}(\mathbf{y}_s; \boldsymbol{\mu}_s,\boldsymbol{\Omega}_s). \label{likelihood}
\end{equation}

Assuming an uninformative prior for the source location, the prior probability $p\left(\mathbf{x}^* \right)$ is specified using a uniform distribution as:
\begin{equation}p(\mathbf{x}^*) = \frac{1}{N_z} \times \frac{1}{N_c} \times \frac{1}{A_z^n}  \label{prior} \end{equation}
where $A_z$ is the area of $z^{th}$ zone.

The unnormalized posterior probability of given contaminant source location can be estimated by using (\ref{bayes}, \ref{likelihood})
\begin{align}p(\mathbf{x}^*|\mathbf{Y}) &\propto p(\mathbf{Y}|\mathbf{x}^*)p(\mathbf{x}^*) \nonumber \\
&= p(\mathbf{x}^*)p(\mathbf{Y}|T(\mathbf{x}^*)\nonumber \\
&= p(\mathbf{x}^*) \prod_{s\in Z_s} \mathcal{N}(\mathbf{y}_s; \boldsymbol{\mu}_s,\boldsymbol{\Omega}_s) , 
\end{align}
while the posterior probability distribution of source locations can be inferred through MCMC sampling method with Metropolis-Hastings algorithm \cite{hastings1970monte}.
The initial MCMC chain is chosen with the maximum a posteriori probability (MAP) point after random sampling. 
The resultant Bayesian framework for source localization is summarized in Algorithm \ref{MCMC}. 

\begin{algorithm}
	\caption{MCMC for Source Localization}
	\label{MCMC}
	\begin{algorithmic}[1]
		\REQUIRE $\mathbf{Y}$: transformed sensor measurements, $Z_s$: sensor locations, $T_o$: observation time instances 
		\STATE Initialize the chain $\mathbf{x}^{(0)}$ with arbitrary locations of sources. Normalize $\mathbf{x}$ from -1 to 1.
		\FOR {$k = 1$ to $no\_init$}
		\STATE Select a new random design point, $\mathbf{x'}$.
		\STATE Compare $p(\mathbf{x'}|\mathbf{Y})$ with $p(\mathbf{x}^{(0)}|\mathbf{Y})$.
		\STATE Update $\mathbf{x}^{(0)}$ with the chain with higher unnormalized posterior.
		\ENDFOR
		\FOR {$k = 1$ to $no\_samples$}
		\STATE Define cardinality of the chain: $m \leftarrow |\mathbf{x}^{(0)}|$.
		\STATE Set perturbation scale: $\eta \in [0,1]$.
		\STATE Sample a random number $\mathbf{u} \in \eta[-1,1]^m$.
		\STATE $\mathbf{x}^{*} \leftarrow \mathbf{x}^{(k-1)} + \mathbf{u}$
		\FOR {$s = Z_s$}
		\STATE Run the DMGPE and predict the contaminant concentration at $s$, $\mathbf{x}^*$, and $\mathcal{T}_o$.
		\ENDFOR
		\STATE Calculate acceptance probability: \newline $a \leftarrow \min\{1, \frac{p(\mathbf{x}^{*}|\mathbf{Y})}{p(\mathbf{x}^{(k-1)}|\mathbf{Y})}\}$
		\STATE Sample a random number $r \in [-1,1]$.
		\IF {$r \le a$}
		\STATE $\mathbf{x}^{(k)} \leftarrow \mathbf{x}^{*}$ // accept
		\ELSE
		\STATE $\mathbf{x}^{(k)} \leftarrow \mathbf{x}^{(k-1)}$ // refuse
		\ENDIF
		\ENDFOR
	\end{algorithmic} 
\end{algorithm}

\subsection{Efficient Prior Information Acquisition}
A key advantage of the Bayesian framework is its ability to admit the prior information together with the observed data.
In this context, it is worthwhile to acquire strong prior information rather than using a uniform prior described in \eqref{prior}. 
Fortunately, identification of the zone with active sources is a standard classification problem that can be solved using modern deep learning algorithms with high accurcay~\cite{bengio2015deep}.
As such, predicting the zone including active sources is a relatively simple classification task that many of recent deep models can accurately perform. Possible candidate models are deep neural networks (DNN), convolutional neural network (CNN), deep belief network (DBN), deep Boltzmann machines (DBM), and so on. Among the model, we utilized DBM since it is known to deal with uncertain and ambiguous data robustly \cite{salakhutdinov2009deep}. See Appendix D for summary of DBM.

In the contaminant localization problem, input data for DBM is whole sensor measurement data $\{Y_j; j=1,\cdots,N_s\}$ while target output is the probability distribution about the zone with active sources.
In the Bayesian framework, DBM output can be used to sharpen the prior of $p(z)$ rather than to describe it by the uniform prior.
Strong prior knowledge not only make the posterior distribution sharp but increase the efficiency of MCMC sampling methods.
To avoid overestimated prior from DBM, this paper guaranteed minimum probability $\epsilon_z = 0.01$ for each zone as prior to make MCMC samples to explore every zones.

Overall flow diagram integrating DMGPE, DBM, and DMGP forecaster is shown in Fig. \ref{architecture}. The role and implementation of DMGP forecaster are described in the later chapter.

\section{Results and Discussion}  \label{sec:results}
\subsection{Environment Settings}
\begin{figure}[t!]
	\centering
	\includegraphics[width=0.5\columnwidth]{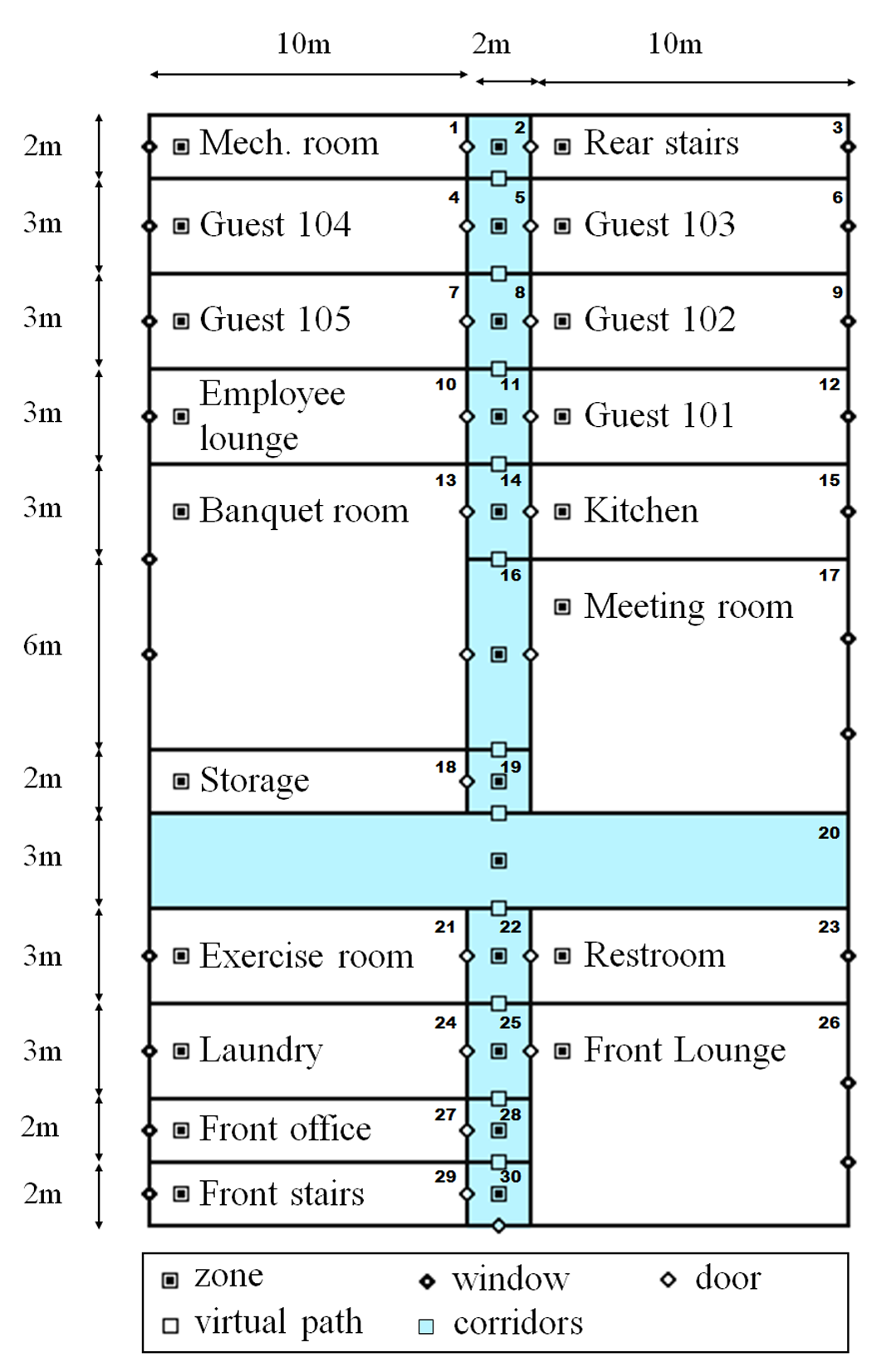}
	\caption{A Building Design}
	\label{building}
\end{figure}

Efficacy of the proposed method is demonstrated for multiple source localization in a hotel building consisting of 30 zones (18 rooms and 12 connected corridors) shown in Fig. \ref{building}. 
Each room is 3.35 m high, has one or more 1 m$\times$1 m windows, and is connected to an adjacent corridor through one or more 1 m$\times$2 m doors (width $\times$ height).
Every window and door is assumed to be open and located middle of the wall. Windows locate 1 m away from floors while doors attached to floors.
Corridors are connected with each other through large virtual path (i.e. walls between corridors have area of 0 m$^2$). The outdoor condition is at 23 \textcelsius , and the wind is assumed to blow at a speed of 3.6 m/s. 
Indoor contaminant release is simulated using the multizone-CFD model, and the output concentration is treated as sensor measurements.
The type of contaminant source is carbon monoxide (CO) and it is released at a rate of 0.2 g/s for 5 minutes from 0 minute.

The zone with active sources is selected as CFD zone whereas other zones are considered as multizone models during the simulation. A single or multiple sources are released in the arbitrary room, and transient contaminant concentration in each zone is obtained by using the multizone-CFD CONTAM simulation.

\begin{figure}[t!]
	\centering
	\includegraphics[width=\columnwidth]{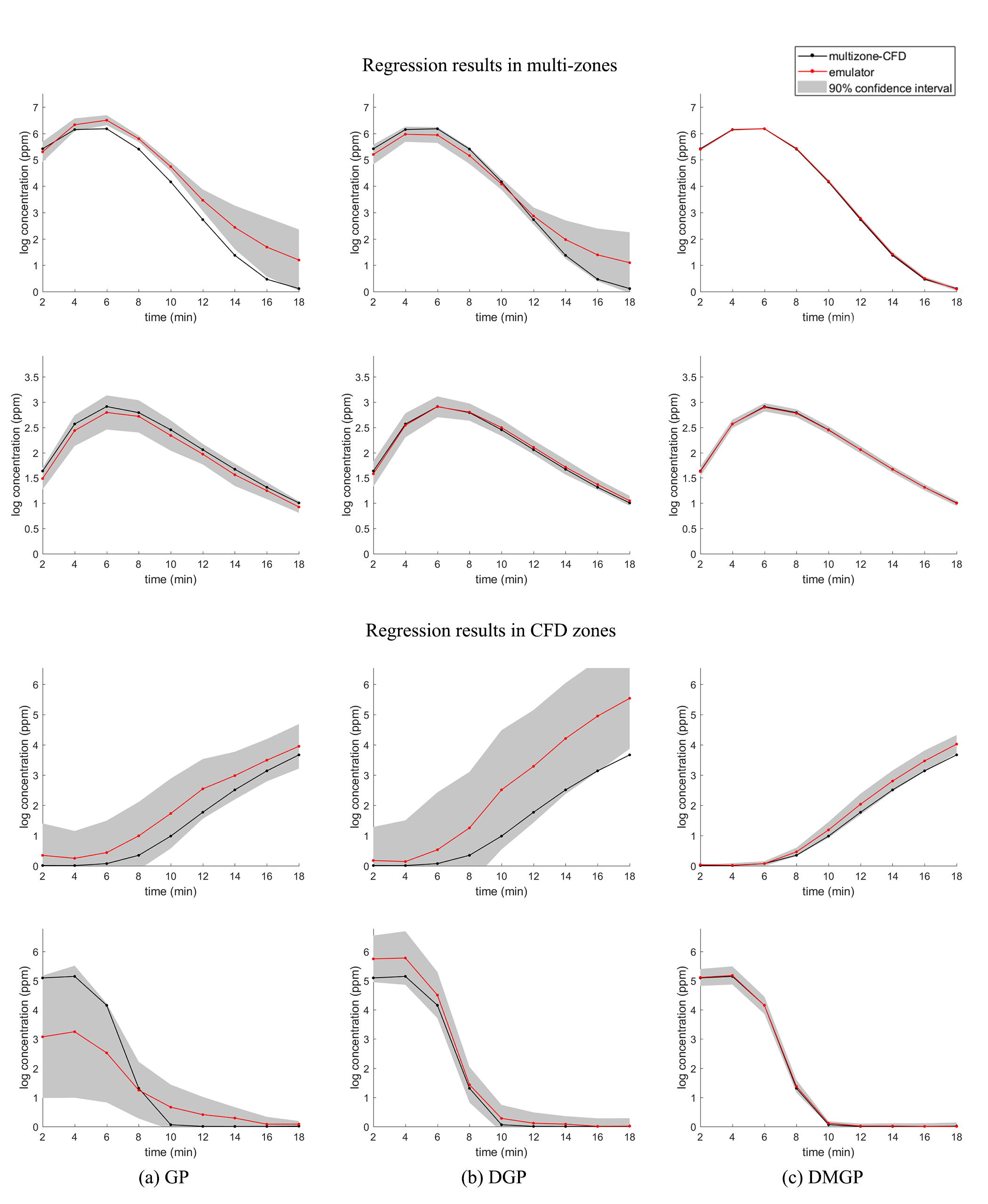}
	\caption{Comparison among GP, DGP, and DMGP emulators at four different design points.}
	\label{resultemul}
\end{figure}

\subsection{Emulator Performance}
To build emulators, training data of 105 to 399 simulation results are used for each emulator according to the size of each room. Each training data is consisted of total 9 measurements data at $\mathcal{T}_o = \{2, 4, 6, 8, 10, 12, 14, 16, 18\}$ minutes. 
DMGP inside the emulator is comprised of 2-layer GPs with 20 inducing variables.
Mean function for each layer has been selected as linear model: $m^l(\mathbf{X}) = \mathbf{X}\mathbf{H}^l+\mathbf{b}^l$.
The number of samples was selected as: $K=10$ for the training and $K=50$ for the prediction.

The performance of DMGPEs is compared with GPEs, DGPEs, and multizone-CFD simulator.
Fig. \ref{resultemul} compares the emulator performance at four different design points: $(z,s)=(1,2)$, $(z,s)=(13,15)$, $(z,s)=(18,18)$, and $(z,s)=(27,27)$.
The respective RMSE and log-likelihood are summarized in Table \ref{resultemul_table}.

First of all, DMGPEs show better performance in terms of both RMSE and log-likelihood compare to other approaches.
Furthermore, DMGPEs show a higher capability to determine the proper level of variances for each time point whereas other models maintain relatively similar variance level among the time domain due to the assumption of probabilistic independence (i.e. the importance of data for each dimension is regarded as equivalent).
As such, DMGPEs try to determine the uncertainty level of emulation adaptively for each dimension (i.e. time point).
Proposed emulators took only 0.1 to 0.4 seconds for the prediction, while the multizone-CFD simulations took about 10 to 50 seconds by using Intel Core i7 4.20GHz CPU, .
	This significant computational speedup allows us to use the proposed framework for online real-time source localization.

\begin{table} [t!]
	\centering
	\caption{RMSE and Log-likelihood of Test Cases in Fig. \ref{resultemul}}
	\label{resultemul_table}
	\begin{tabular}{|c|ccc|ccc|}
		\hline
		& \multicolumn{3}{c|}{RMSE} & \multicolumn{3}{c|}{log-likelihood} \\ 
		$(z,s)$ & GPE & DGPE & DGMPE & GPE & DGPE & DGMPE \\ \hline
		(1,2)   & 2.230 & 1.549 & \textbf{0.085} & -3936.2 & -3900.9 & \textbf{-3877.9} \\
		(13,15) & 0.319 & 0.121 & \textbf{0.024} & -3891.8 & -3886.9 & \textbf{-3878.8} \\
		(18,18) & 1.523 & 3.928 & \textbf{0.668} & -3904.9 & -3913.7 & \textbf{-3893.5} \\
		(27,27) & 3.312 & 1.012 & \textbf{0.094} & -3905.8 & -3898.0 & \textbf{-3886.4} \\  \hline
	\end{tabular}
\end{table}

\subsection{The MCO and Training Speed}
\begin{figure}[t!]
	\centering
	\includegraphics[width=0.5\columnwidth]{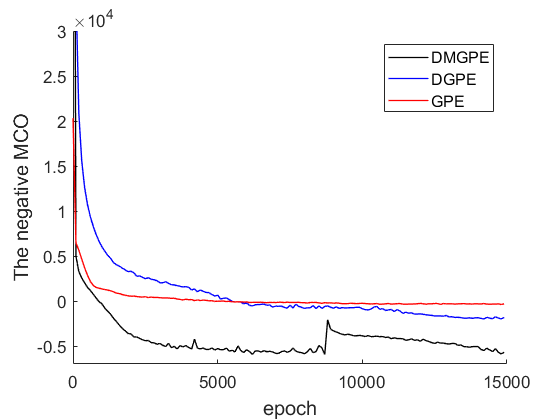}
	\caption{The MCO during a training for each emulator of 17$^{th}$ zone.}
	\label{training}
\end{figure}

As discussed previously, the DMGPE have two significant advantages compared to existing DGPE or GPE.
Primarily, the DMGPE possess higher expressive power by introducing output covariance.
Secondly, DGMPE have approximately $D$ times smaller number of hyperparameters compare to the DGPE that facilitate faster model parameter optimization.
To support above two statements, the MCO and training speed of GPEs, DGPEs, and DMGPEs are studied and compared in Fig. \ref{training} and Table \ref{modelMCO}.
A shown in results, not only the DMGPE reaches to the lowest negative bound, but matrix-variate formulation improves the training speed.
We further observed that time to train DGPEs was about twice longer than DMGPEs for the same epoch length in practice \footnote{Using single-GPU, DMGPEs were trained within 2 to 3 minutes for each zone while DGPEs took about 5 to 6 minutes.}.

\begin{table}[t!]
	\centering
	\caption{The Lower Bound of Trained GPEs, DPGEs, and DMGPEs.}
	\label{modelMCO}
	\begin{tabular}{|c|ccc|}
		\hline
		\multicolumn{1}{|c|}{\multirow{2}{*}{Active Zone}} & \multicolumn{3}{c|}{The Lower Bound} \\
		\multicolumn{1}{|c|}{} & GPE    & DGPE   & DMGPE  \\ \hline
		1  & -386.7  & -428.6 & \textbf{1787.0} \\
		3  & -231.4  & 76.7   & \textbf{1643.9} \\
		4  & -502.1  & 449.6  & \textbf{1957.1} \\
		6  & -296.4  & 440.5  & \textbf{2595.0} \\
		7  & -458.4  & -758.6 & \textbf{1843.2} \\
		9  & -796.0  & -47.9  & \textbf{1800.4} \\
		10 & -447.4  & -42.7  & \textbf{1752.7} \\
		12 & -596.5  & 375.0  & \textbf{1760.8} \\
		13 & 4384.8  & 3045.4 & \textbf{8406.4} \\
		15 & -365.1  & 409.7  & \textbf{2142.7} \\
		17 & -680.6  & 3057.3 & \textbf{5933.9} \\
		18 & -1002.6 & -25.0  & \textbf{1336.9} \\
		21 & -199.5  & 610.3  & \textbf{2259.0} \\
		23 & -309.5  & 217.1  & \textbf{2412.0} \\
		24 & -1078.5 & 262.9  & \textbf{2693.1} \\
		26 & 96.0    & 1794.5 & \textbf{6179.9} \\
		27 & -60.6   & 327.0  & \textbf{1363.2} \\
		29 & -152.2  & 188.8  & \textbf{1472.2} \\ \hline
	\end{tabular}
\end{table}

\begin{figure}[t!]
	\includegraphics[width=\columnwidth]{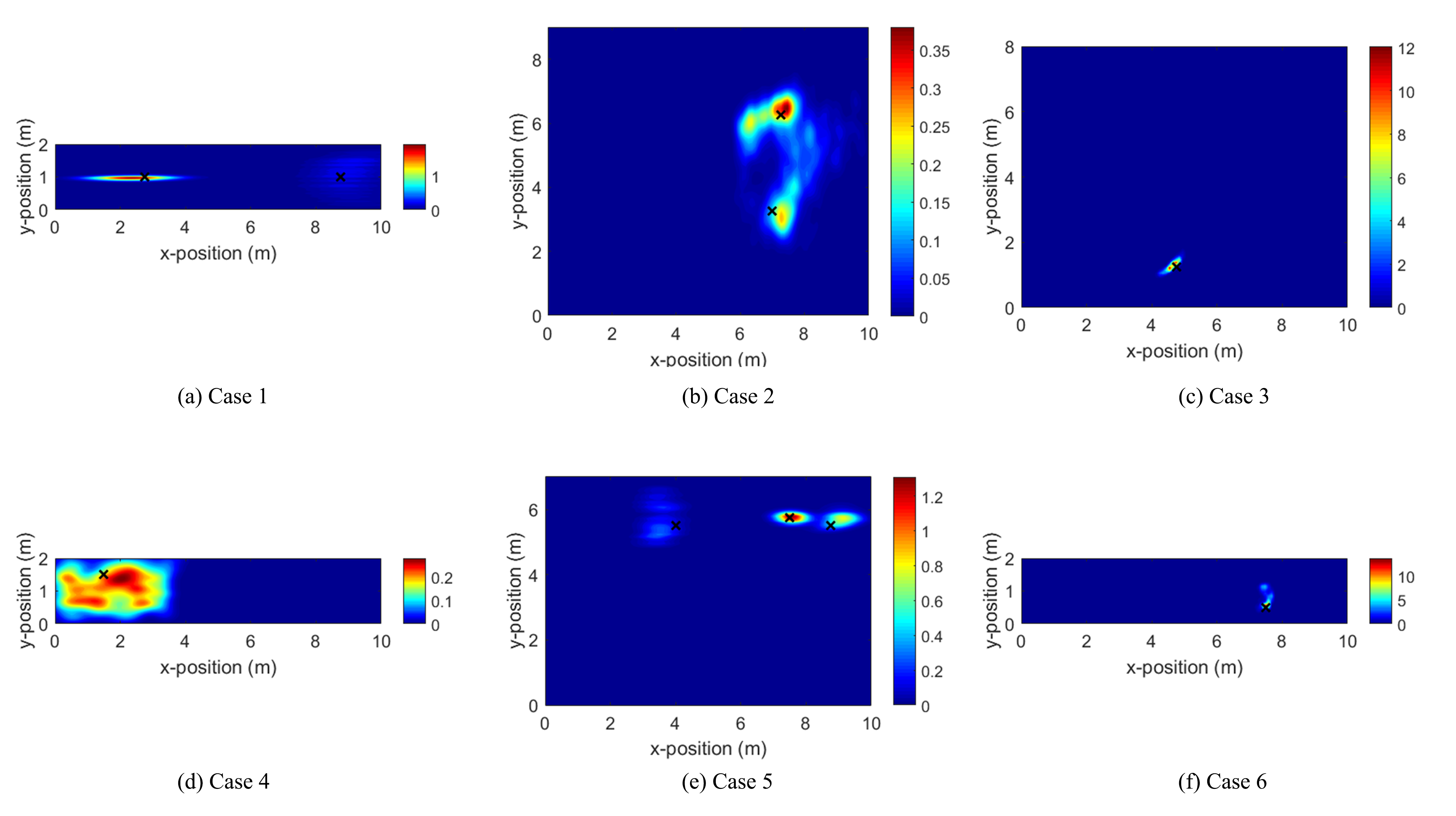}
	\caption{Posterior probability distribution of contaminant locations. Color represents the probability density, and true sources are illustrated by black x markers.}
	\label{resultloc}
\end{figure}

\subsection{Contaminant Source Localization}
\subsubsection{Test Cases}
A building with 18 sensors measuring average contaminant concentration in each room is considered.
For each sensor, 7 transient concentration data from 2 to 14 minutes are used for the inference.
Depends on the source location, some sensors may not be able to detect any contaminant if the source locates too far away from the sensor or flows outside through windows.
However, the distribution of active and inactive sensors also provide important information to distinguish the zone including active sources.

Six test cases described in Table \ref{cases} are concerned in this paper. We simulate circumstances that pollutants release CO gas in mechanical room (1$^{st}$ zone), banquet room (13$^{th}$ zone), meeting room (17$^{th}$ zone), storage (18$^{th}$ zone), front lounge (26$^{th}$ zone), and front office (27$^{th}$ zone). Case 3, 4 and 6 are dealing with single source cases, while case 1, 2, and 5 consider multiple sources cases. 

Contaminant source localization is done in two steps.
Firstly, prior information is acquired by using DBM.
In the next step, the inference of contaminant source locations has been performed by following Algorithm \ref{MCMC}.
Perturbation scale $\eta$ was set as $0.1$.

\begin{table} [t!]
	\centering
	\caption{Test Cases Descriptions.}
	\label{cases}
	\begin{tabular}{|c|c|c|c|}
		\hline
		~ Case ~ & ~ $z_{true}$ ~ & $n_{true}$& $\mathbf{x}_{true}$ (m)\\
		\hline
		1 & 1 & 2 & (2.75, 1.00), (8.75, 1.00) \\
		2 & 13 & 2 & (7.25, 6.25), (7.00, 3.25) \\
		3 & 17 & 1 & (4.75, 1.25) \\
		4 & 18 & 1 & (1.50, 1.50) \\
		5 & 26 & 3 & (4.00, 5.50), (7.50, 5.75), (8.75, 5.50) \\
		6 & 27 & 1 & (7.50, 0.50) \\
		\hline
		\multicolumn{4}{p{230pt}}{$z_{true}$ : Zone including active sources. ~ $n_{true}$ : The number of active sources. ~ $\mathbf{x}_{true}$ : Locations of active sources} 
	\end{tabular}
\end{table}

\subsubsection{Source Localization Results}
Fig. \ref{resultloc} shows the posterior probability distribution map of contaminant locations inferred by MCMC sampling. Table \ref{prob} summarizes location error and the prior, posterior probability of contaminant characteristics. Results show that DMGPE based MCMC algorithm accurately the infers the contaminant locations for both single and multiple sources cases. Appendix E further shows that the DMGPE based framework outperforms GPE and DGPE based frameworks. Average absolute distance error is less than 0.6 meters. DMGPE framework also accurately estimates the contaminant characteristics. Single source cases show higher estimation accuracy than multiple sources case in general. This is due to the fact that emulation error and regression uncertainty proportionally increases to the number of sources.

\begin{table} [t!]
	\centering
	\caption{Contaminant Localization Results.}
	\label{prob}
	\begin{tabular}{|c|cc|cc|c|}
		\hline
		& \multicolumn{2}{c|}{Prior (DBM)} & \multicolumn{2}{c|}{Posterior} & Average location error (m)\\ 
		~~~ Case ~~~ & $p(z)$ & $p(n)$ & $p(z)$ & $p(n)$ & $\sum_{i=1}^n||\mathbf{x}-\mathbf{x}_{true}||/n$ \\ \hline
		1 &  0.83 & 0.33  & 1.00  & 1.00 & 0.31 \\
		2 &  0.83 & 0.33  & 1.00  & 1.00 & 0.56 \\
		3 &  0.83 & 0.33  & 1.00  & 0.93 & 0.13 \\
		4 &  0.83 & 0.33  & 1.00  & 0.93 & 0.48 \\
		5 &  0.83 & 0.33  & 1.00  & 1.00 & 0.31 \\
		6 &  0.83 & 0.33  & 1.00  & 0.74 & 0.26 \\
		\hline
		\multicolumn{6}{p{230pt}}{$p(z)$ and $p(n)$ represents $p(z=z_{true})$ and $p(n=n_{true})$, respectively.}
	\end{tabular}
\end{table}

\subsubsection{Efficient Sampling via DBM}
\begin{figure}[t!]
	\centering
	\includegraphics[width=0.5\columnwidth]{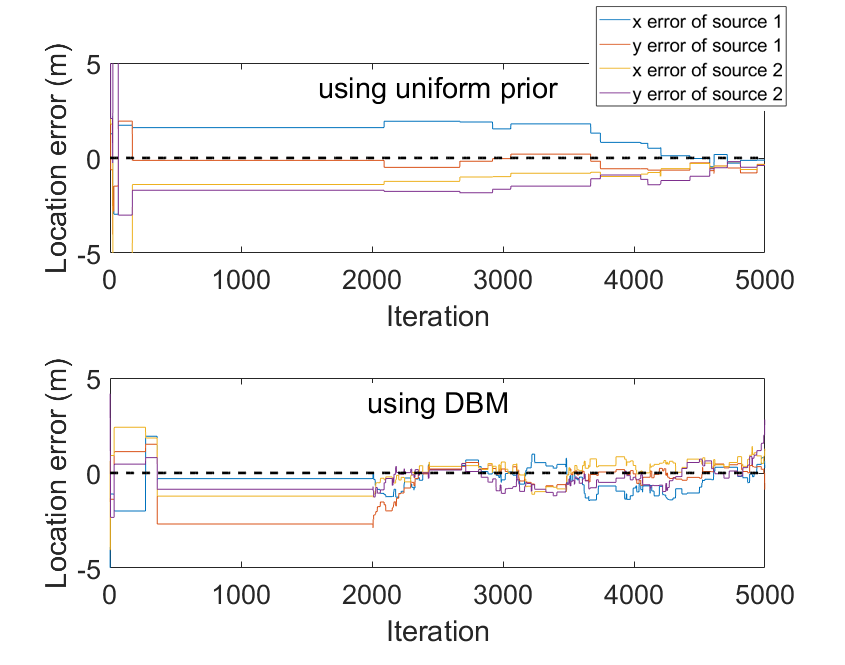}
	\caption{Location error plots using uniform prior and DBM.}
	\label{dbm_plot}
\end{figure}

Table \ref{prob} shows that the DBM gives very sharp and accurate prior distribution about the zone with active sources, and we further investigate the effect of DBM on MCMC sampling efficiency.
The result of case 2 in Table \ref{cases} is compared with the one using the uniform prior illustrated in \eqref{prior}.
Fig. \ref{dbm_plot} shows location error during 5000 iterations ($no\_init=2000$ and $no\_samples=3000$) for both cases.
A result coincides with our intuition that convergence speed becomes much faster by using DBM; MCMC samples converge within 500 iterations from the initial MCMC chain with DBM while the uniform prior sample very slowly moves toward true locations. 
Thanks to the strong prior, MCMC samples can focus on the most probable zone rather than searching the whole building, and the number of MCMC samples could be effectively reduced by the number of zones.

Note that we can easily specify the active zone at high accuracy even with information about the state of activeness in each room.
As such, the zone classification task seems relatively easier compared to emulations in contaminant problems.
Empirically, we found that DBM gives higher than 99\% accuracy for the zone classification tasks.

\subsubsection{Effect of Varying Perturbation Scale}
\begin{figure}[t!]
	\centering
	\includegraphics[width=0.5\columnwidth]{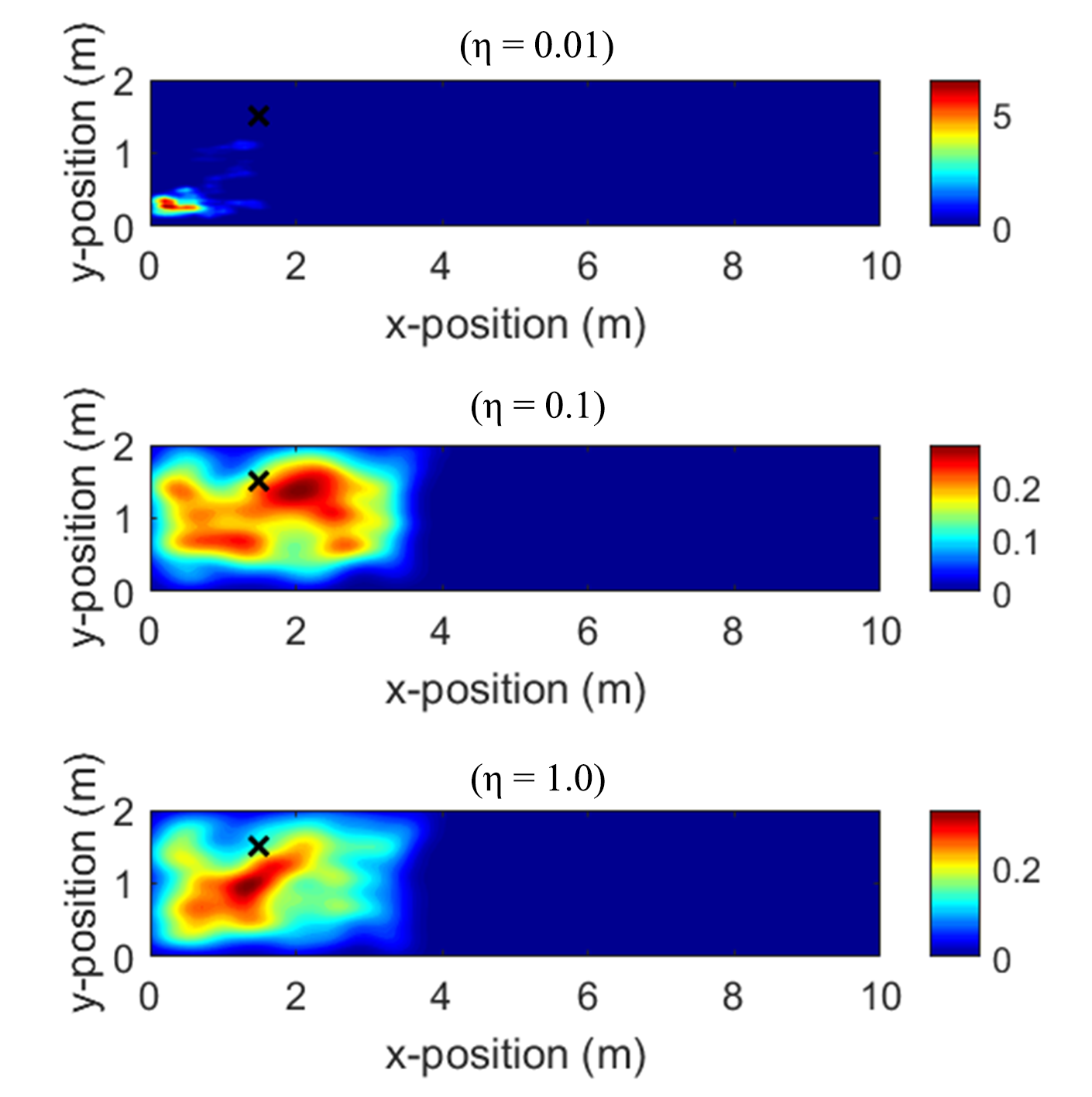}
	\caption{Contaminant sources localization results by perturbation scale.}
	\label{fig:perturb}
\end{figure}

As illustrated line 15-21 in Algorithm \ref{MCMC}, if the MCMC sample locates too far from posterior distribution, acceptance probability will be too low and it is highly probable to be refused.
Thus it is important to select a proper size of perturbation scale $\eta$ for better sample efficiency.
To investigate the effect of perturbation scale to source localization result quality, we experimented with three different perturbation scale, $\eta=0.01, 0.1, 1.0$, with the test case 4 in Table \ref{cases}.
The posterior distribution of localized contaminant source along time is shown in Fig. \ref{fig:perturb}.
Acceptance rate for each perturbation scale is observed as $40 \pm 2 \%$, $32 \pm 1 \%$, and $17 \pm 0.5 \%$ for $\eta=0.01, 0.1, 1.0$, respectively.
As can be observed from the results, too small perturbation scale disturbs the sample exploration and yield poor results.
Meanwhile too large perturbation scale lowers the acceptance rate, that leads to reduction in the sample efficiency. 
Empirically is found that $\eta=0.1$ is suitable scale for contaminant source localization problems.

\subsubsection{Effect of Varying Observation Length}
\begin{figure}[t!]
	\centering
	\includegraphics[width=0.5\columnwidth]{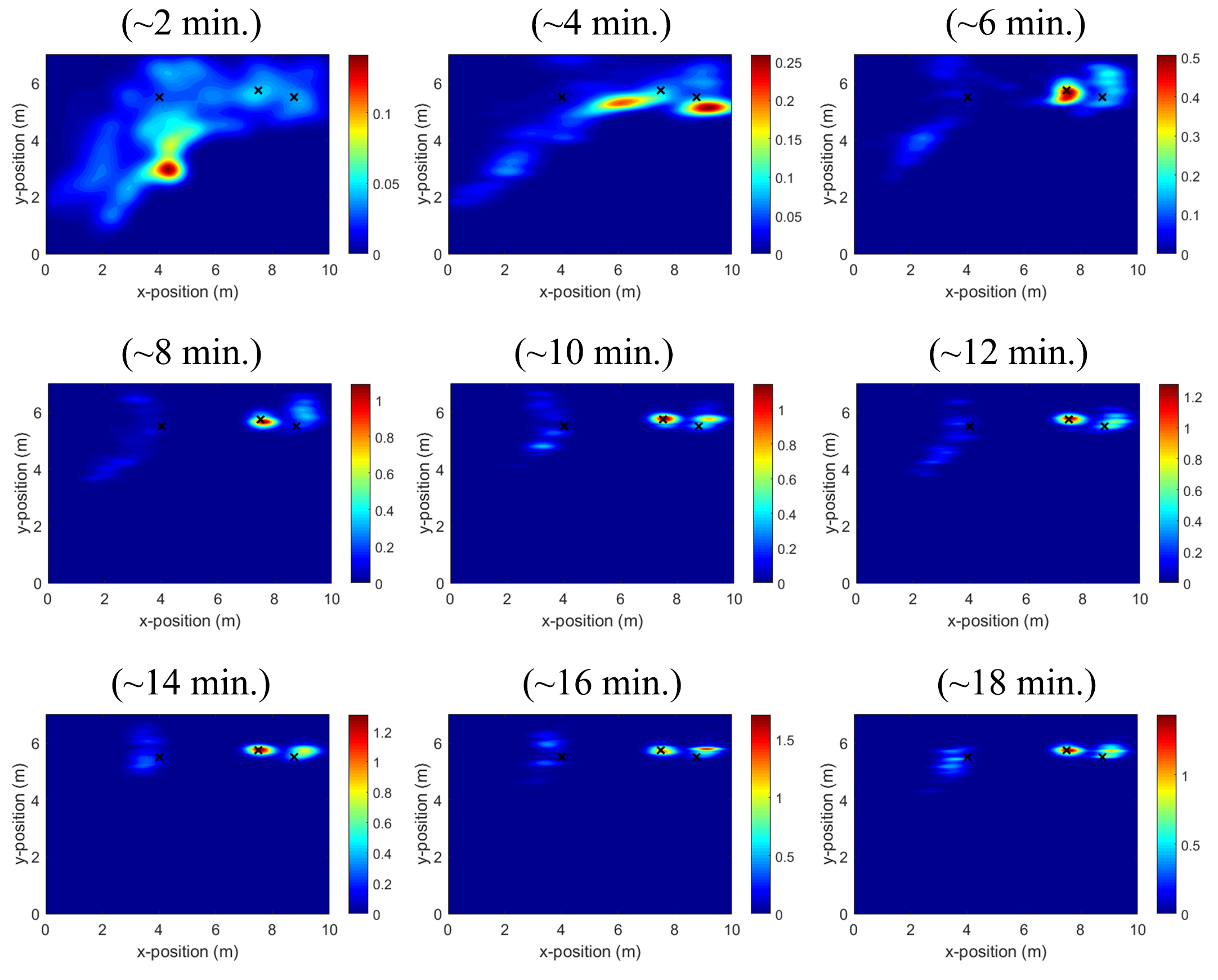}
	\caption{Contaminant sources localization results by observation time.}
	\label{time_mc}
\end{figure}

\begin{figure}[t!]
	\centering
	\includegraphics[width=0.5\columnwidth]{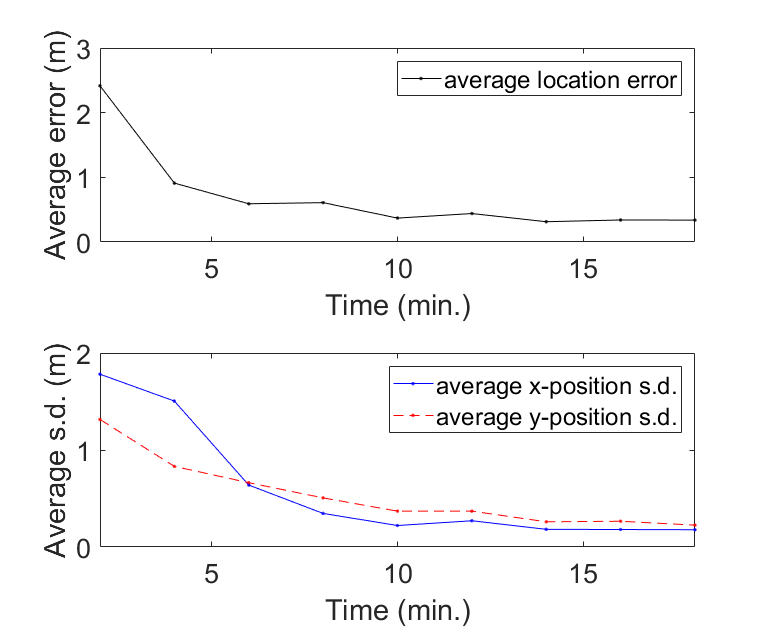}
	\caption{Average location error and standard deviation by observation time.}
	\label{time_plot}
\end{figure}

Finally, the performance of proposed framework with different observation length is investigated.
Too short sensor data can not provide an unique and accurate solution while too long sensor data is excessive use of time.
Case 5 in Table \ref{cases} has been investigated, where we obtain an optimum amount of data necessary to detect and localize multiple sources to prepare emergency.
Fig. \ref{time_mc} and Fig. \ref{time_plot} shows that localization becomes more accurate and precise as observation time gets longer.
Empirically is found that 14 minutes is an optimum length as the localization accuracy converges approximately after 14 minutes.

\subsection{Prediction of Future Contaminant Transport}
\begin{figure}[t!]
	\centering
	\includegraphics[width=0.5\columnwidth]{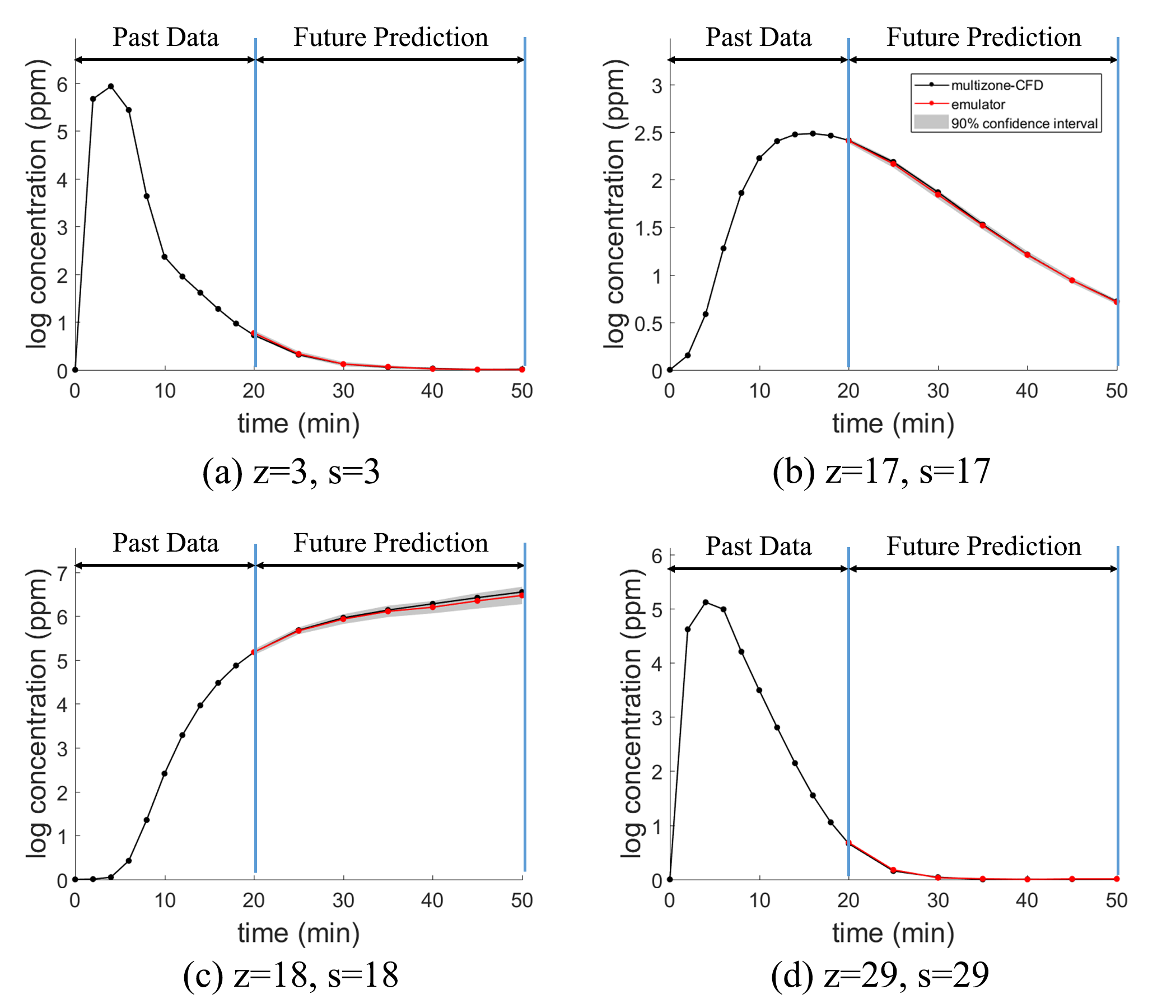}
	\caption{Future prediction by using DMGP forecaster.}
	\label{forecast}
\end{figure}

As well as determining contaminant sources locations, predicting the future transport of target contaminant is an important task for efficient management of HVAC systems. Future contaminant transport path can be determined from the past contaminant dispersion pattern.
In case of CFD simulation, states of contaminant evolve via partial differential equation of fluid dynamics described in \eqref{CFD} that depends on the past state.
So it is reasonable to expect past sensor data can be used for the long-term time series forecasting \cite{roberts2013gaussian, richardson2017gaussian, rogers2011adaptive}.
In this point of view, we build DMGP forecasters (DMGPFs) that get input of past sensor measurements $\mathbf{Y}_j$ and generate output of future contaminant concentration, $C_j(\mathcal{T}_{future}; \gamma, \mathbf{X})$.
As we have built emulators, DMGP inside the forecaster is comprised of 2-layer GPs with 20 inducing variables.
Mean function for each layer has been selected as linear model: $m^l(\mathbf{X}) = \mathbf{X}\mathbf{H}^l+\mathbf{b}^l$.
The number of samples was selected as: $K=10$ for the training and $K=50$ for the prediction.
Fig. \ref{forecast} supports the applicability of DMGP to forecaster providing an accurate future prediction with low variance.
Further results comparing DMGPF with DGP forecaster (DGPF) and GP forecaster (GPF) are presented in Appendix E.

\section{Conclusion}\label{sec:con}
In this paper, we presented a deep matrix-variate Gaussian process emulator based Bayesian framework to perform rapid and accurate multiple contaminant sources localizations. To take account for correlation over time domain, DGP was extended to DMGP. To train the DMGP model, the stochastic variational inference method has been used to compute the lower bound of the log marginal likelihood of the model. Based on DMGP, emulators were built to approximate computationally heavy integrated multizone-CFD simulator in real-time. Implemented DMGPEs aided MCMC algorithm to rapidly and accurately perform Bayesian inference of source locations in a 30 zone building. Single and multiple sources cases were studied, and proposed methods well inferred the zone including active sources, the number of sources, and the posterior probability distribution of sources locations within 0.6 m error. Furthermore, the study illustrated efficient method to acquire strong prior about the zone with active sources by using deep Boltzmann machine. In future, the proposed method can be extended to more complex cases like multiple story entire CFD zone building. Another interesting topic is to develop sensor planning algorithm by using DMGP forecaster.

\section*{Acknowledgments}
The first and the third author thank financial support from the KAIST grant EEWS-2017-N1170058.

\bibliographystyle{IEEEtran}
\bibliography{manuscript}

\begin{thebibliography}{10}
\providecommand{\url}[1]{#1}
\csname url@samestyle\endcsname
\providecommand{\newblock}{\relax}
\providecommand{\bibinfo}[2]{#2}
\providecommand{\BIBentrySTDinterwordspacing}{\spaceskip=0pt\relax}
\providecommand{\BIBentryALTinterwordstretchfactor}{4}
\providecommand{\BIBentryALTinterwordspacing}{\spaceskip=\fontdimen2\font plus
\BIBentryALTinterwordstretchfactor\fontdimen3\font minus
  \fontdimen4\font\relax}
\providecommand{\BIBforeignlanguage}[2]{{%
\expandafter\ifx\csname l@#1\endcsname\relax
\typeout{** WARNING: IEEEtran.bst: No hyphenation pattern has been}%
\typeout{** loaded for the language `#1'. Using the pattern for}%
\typeout{** the default language instead.}%
\else
\language=\csname l@#1\endcsname
\fi
#2}}
\providecommand{\BIBdecl}{\relax}
\BIBdecl

\bibitem{abraham2014cost}
S.~Abraham and X.~Li, ``A cost-effective wireless sensor network system for
  indoor air quality monitoring applications,'' \emph{Procedia Computer
  Science}, vol.~34, pp. 165--171, 2014.

\bibitem{kim2014issaq}
J.-Y. Kim, C.-H. Chu, and S.-M. Shin, ``Issaq: An integrated sensing systems
  for real-time indoor air quality monitoring,'' \emph{IEEE Sensors Journal},
  vol.~14, no.~12, pp. 4230--4244, 2014.

\bibitem{chen2014indoor}
X.~Chen, Y.~Zheng, Y.~Chen, Q.~Jin, W.~Sun, E.~Chang, and W.-Y. Ma, ``Indoor
  air quality monitoring system for smart buildings,'' in \emph{Proceedings of
  the 2014 ACM International Joint Conference on Pervasive and Ubiquitous
  Computing}.\hskip 1em plus 0.5em minus 0.4em\relax ACM, 2014, pp. 471--475.

\bibitem{kumar2016indoor}
P.~Kumar, C.~Martani, L.~Morawska, L.~Norford, R.~Choudhary, M.~Bell, and
  M.~Leach, ``Indoor air quality and energy management through real-time
  sensing in commercial buildings,'' \emph{Energy and Buildings}, vol. 111, pp.
  145--153, 2016.

\bibitem{kennedy2001bayesian}
M.~C. Kennedy and A.~O'Hagan, ``Bayesian calibration of computer models,''
  \emph{Journal of the Royal Statistical Society: Series B (Statistical
  Methodology)}, vol.~63, no.~3, pp. 425--464, 2001.

\bibitem{tagade2009bayesian}
P.~M. Tagade, K.~Sudhakar, and S.~K. Sane, ``Bayesian framework for calibration
  of gas turbine simulator,'' \emph{Journal of Propulsion and Power}, vol.~25,
  no.~4, pp. 987--992, 2009.

\bibitem{spengler1983indoor}
J.~D. Spengler and K.~Sexton, ``Indoor air pollution: a public health
  perspective,'' \emph{Science}, vol. 221, no. 4605, pp. 9--17, 1983.

\bibitem{jones1999indoor}
A.~P. Jones, ``Indoor air quality and health,'' \emph{Atmospheric environment},
  vol.~33, no.~28, pp. 4535--4564, 1999.

\bibitem{raub2000carbon}
J.~A. Raub, M.~Mathieu-Nolf, N.~B. Hampson, and S.~R. Thom, ``Carbon monoxide
  poisoning—a public health perspective,'' \emph{Toxicology}, vol. 145,
  no.~1, pp. 1--14, 2000.

\bibitem{chen2008sensor}
Y.~L. Chen and J.~Wen, ``Sensor system design for building indoor air
  protection,'' \emph{Building and Environment}, vol.~43, no.~7, pp.
  1278--1285, 2008.

\bibitem{chen2010comparison}
------, ``Comparison of sensor systems designed using multizone, zonal, and cfd
  data for protection of indoor environments,'' \emph{Building and
  Environment}, vol.~45, no.~4, pp. 1061--1071, 2010.

\bibitem{mahar1997optimal}
P.~S. Mahar and B.~Datta, ``Optimal monitoring network and
  ground-water--pollution source identification,'' \emph{Journal of water
  resources planning and management}, vol. 123, no.~4, pp. 199--207, 1997.

\bibitem{federspiel1997estimating}
C.~C. Federspiel, ``Estimating the inputs of gas transport processes in
  buildings,'' \emph{IEEE Transactions on Control Systems Technology}, vol.~5,
  no.~5, pp. 480--489, 1997.

\bibitem{liu2007inverse}
X.~Liu and Z.~Zhai, ``Inverse modeling methods for indoor airborne pollutant
  tracking: literature review and fundamentals,'' \emph{Indoor air}, vol.~17,
  no.~6, pp. 419--438, 2007.

\bibitem{zhang2007identification}
T.~Zhang and Q.~Chen, ``Identification of contaminant sources in enclosed
  environments by inverse cfd modeling,'' \emph{Indoor air}, vol.~17, no.~3,
  pp. 167--177, 2007.

\bibitem{keats2007bayesian}
A.~Keats, E.~Yee, and F.-S. Lien, ``Bayesian inference for source determination
  with applications to a complex urban environment,'' \emph{Atmospheric
  environment}, vol.~41, no.~3, pp. 465--479, 2007.

\bibitem{sreedharan2007bayesian}
P.~Sreedharan, \emph{Bayesian based design of real-time sensor systems for
  high-risk indoor contaminants}.\hskip 1em plus 0.5em minus 0.4em\relax
  University of California, Berkeley, 2007.

\bibitem{sohn2002rapidly}
M.~D. Sohn, P.~Reynolds, N.~Singh, and A.~J. Gadgil, ``Rapidly locating and
  characterizing pollutant releases in buildings,'' \emph{Journal of the Air \&
  Waste Management Association}, vol.~52, no.~12, pp. 1422--1432, 2002.

\bibitem{tagade2013gaussian}
P.~M. Tagade, B.-M. Jeong, and H.-L. Choi, ``A gaussian process emulator
  approach for rapid contaminant characterization with an integrated
  multizone-cfd model,'' \emph{Building and EnvironmenfGt}, vol.~70, pp.
  232--244, 2013.

\bibitem{hastings1970monte}
W.~K. Hastings, ``Monte carlo sampling methods using markov chains and their
  applications,'' \emph{Biometrika}, vol.~57, no.~1, pp. 97--109, 1970.

\bibitem{borysiewicz2012bayesian}
M.~Borysiewicz, A.~Wawrzynczak, and P.~Kopka, ``Bayesian-based methods for the
  estimation of the unknown model’s parameters in the case of the
  localization of the atmospheric contamination source,'' \emph{Foundations of
  Computing and Decision Sciences}, vol.~37, no.~4, pp. 253--270, 2012.

\bibitem{wawrzynczak2013sequential}
A.~Wawrzynczak, P.~Kopka, and M.~Borysiewicz, ``Sequential monte carlo in
  bayesian assessment of contaminant source localization based on the sensors
  concentration measurements,'' in \emph{International Conference on Parallel
  Processing and Applied Mathematics}.\hskip 1em plus 0.5em minus 0.4em\relax
  Springer, 2013, pp. 407--417.

\bibitem{conti2010bayesian}
S.~Conti and A.~O’Hagan, ``Bayesian emulation of complex multi-output and
  dynamic computer models,'' \emph{Journal of statistical planning and
  inference}, vol. 140, no.~3, pp. 640--651, 2010.

\bibitem{manic2016intelligent}
M.~Manic, K.~Amarasinghe, J.~J. Rodriguez-Andina, and C.~Rieger, ``Intelligent
  buildings of the future: Cyberaware, deep learning powered, and human
  interacting,'' \emph{IEEE Industrial Electronics Magazine}, vol.~10, no.~4,
  pp. 32--49, 2016.

\bibitem{damianou2015deep}
A.~Damianou, ``Deep gaussian processes and variational propagation of
  uncertainty,'' Ph.D. dissertation, University of Sheffield, 2015.

\bibitem{salimbeni2017doubly}
H.~Salimbeni and M.~Deisenroth, ``Doubly stochastic variational inference for
  deep gaussian processes,'' in \emph{Advances in Neural Information Processing
  Systems}, 2017, pp. 4591--4602.

\bibitem{bui2016deep}
T.~Bui, D.~Hern{\'a}ndez-Lobato, J.~Hernandez-Lobato, Y.~Li, and R.~Turner,
  ``Deep gaussian processes for regression using approximate expectation
  propagation,'' in \emph{International Conference on Machine Learning}, 2016,
  pp. 1472--1481.

\bibitem{damianou2013deep}
A.~Damianou and N.~Lawrence, ``Deep gaussian processes,'' in \emph{Artificial
  Intelligence and Statistics}, 2013, pp. 207--215.

\bibitem{boyle2005dependent}
P.~Boyle and M.~Frean, ``Dependent gaussian processes,'' in \emph{Advances in
  neural information processing systems}, 2005, pp. 217--224.

\bibitem{alvarez2009sparse}
M.~Alvarez and N.~D. Lawrence, ``Sparse convolved gaussian processes for
  multi-output regression,'' in \emph{Advances in neural information processing
  systems}, 2009, pp. 57--64.

\bibitem{bonilla2008multi}
E.~V. Bonilla, K.~M. Chai, and C.~Williams, ``Multi-task gaussian process
  prediction,'' in \emph{Advances in neural information processing systems},
  2008, pp. 153--160.

\bibitem{nguyen2014collaborative}
T.~V. Nguyen, E.~V. Bonilla \emph{et~al.}, ``Collaborative multi-output
  gaussian processes.'' in \emph{UAI}, 2014, pp. 643--652.

\bibitem{gupta1999matrix}
A.~K. Gupta and D.~K. Nagar, \emph{Matrix variate distributions}.\hskip 1em
  plus 0.5em minus 0.4em\relax CRC Press, 1999, vol. 104.

\bibitem{tagade2016bayesian}
P.~Tagade, K.~S. Hariharan, S.~Basu, M.~K.~S. Verma, S.~M. Kolake, T.~Song,
  D.~Oh, T.~Yeo, and S.~Doo, ``Bayesian calibration for electrochemical thermal
  model of lithium-ion cells,'' \emph{Journal of Power Sources}, vol. 320, pp.
  296--309, 2016.

\bibitem{burda2015importance}
Y.~Burda, R.~Grosse, and R.~Salakhutdinov, ``Importance weighted
  autoencoders,'' \emph{arXiv preprint arXiv:1509.00519}, 2015.

\bibitem{mnih2016variational}
A.~Mnih and D.~J. Rezende, ``Variational inference for monte carlo
  objectives,'' \emph{arXiv preprint arXiv:1602.06725}, 2016.

\bibitem{louizos2016structured}
C.~Louizos and M.~Welling, ``Structured and efficient variational deep learning
  with matrix gaussian posteriors,'' in \emph{International Conference on
  Machine Learning}, 2016, pp. 1708--1716.

\bibitem{liu2009prompt}
X.~Liu and Z.~J. Zhai, ``Prompt tracking of indoor airborne contaminant source
  location with probability-based inverse multi-zone modeling,'' \emph{Building
  and Environment}, vol.~44, no.~6, pp. 1135--1143, 2009.

\bibitem{wang2007coupling}
L.~Wang, ``Coupling of multizone and cfd programs for building airflow and
  contaminant transport simulations,'' Ph.D. dissertation, Purdue University,
  2007.

\bibitem{tan2005application}
G.~Tan and L.~R. Glicksman, ``Application of integrating multi-zone model with
  cfd simulation to natural ventilation prediction,'' \emph{Energy and
  Buildings}, vol.~37, no.~10, pp. 1049--1057, 2005.

\bibitem{kim2015development}
D.~Kim, J.~Braun, E.~Cliff, and J.~Borggaard, ``Development, validation and
  application of a coupled reduced-order cfd model for building control
  applications,'' \emph{Building and Environment}, vol.~93, pp. 97--111, 2015.

\bibitem{dols2015development}
W.~S. Dols, L.~Wang, S.~J. Emmerich, and B.~J. Polidoro, ``Development and
  application of an updated whole-building coupled thermal, airflow and
  contaminant transport simulation program (trnsys/contam),'' \emph{Journal of
  Building Performance Simulation}, vol.~8, no.~5, pp. 326--337, 2015.

\bibitem{dols2015contam}
W.~S. Dols and B.~J. Polidoro, ``Contam user guide and program documentation
  version 3.2,'' \emph{Technical Note (NIST TN)-1887}, no.~1, 2015.

\bibitem{rasmussen2006gaussian}
C.~E. Rasmussen and C.~K. Williams, \emph{Gaussian processes for machine
  learning}.\hskip 1em plus 0.5em minus 0.4em\relax MIT press Cambridge, 2006,
  vol.~1.

\bibitem{lawrence2007hierarchical}
N.~D. Lawrence and A.~J. Moore, ``Hierarchical gaussian process latent variable
  models,'' in \emph{Proceedings of the 24th international conference on
  Machine learning}.\hskip 1em plus 0.5em minus 0.4em\relax ACM, 2007, pp.
  481--488.

\bibitem{snelson2006sparse}
E.~Snelson and Z.~Ghahramani, ``Sparse gaussian processes using
  pseudo-inputs,'' in \emph{Advances in neural information processing systems},
  2006, pp. 1257--1264.

\bibitem{titsias2009variational}
M.~K. Titsias, ``Variational learning of inducing variables in sparse gaussian
  processes,'' in \emph{International Conference on Artificial Intelligence and
  Statistics}, 2009, pp. 567--574.

\bibitem{hensman2013gaussian}
J.~Hensman, N.~Fusi, and N.~D. Lawrence, ``Gaussian processes for big data,''
  \emph{arXiv preprint arXiv:1309.6835}, 2013.

\bibitem{gal2015latent}
Y.~Gal, Y.~Chen, and Z.~Ghahramani, ``Latent gaussian processes for
  distribution estimation of multivariate categorical data,'' 2015.

\bibitem{bonilla2016generic}
E.~V. Bonilla, K.~Krauth, and A.~Dezfouli, ``Generic inference in latent
  gaussian process models,'' \emph{arXiv preprint arXiv:1609.00577}, 2016.

\bibitem{kingma2013auto}
D.~P. Kingma and M.~Welling, ``Auto-encoding variational bayes,'' \emph{arXiv
  preprint arXiv:1312.6114}, 2013.

\bibitem{kingma2014adam}
D.~Kingma and J.~Ba, ``Adam: A method for stochastic optimization,''
  \emph{arXiv preprint arXiv:1412.6980}, 2014.

\bibitem{abadi2016tensorflow}
M.~Abadi, P.~Barham, J.~Chen, Z.~Chen, A.~Davis, J.~Dean, M.~Devin,
  S.~Ghemawat, G.~Irving, M.~Isard \emph{et~al.}, ``Tensorflow: A system for
  large-scale machine learning.'' in \emph{OSDI}, vol.~16, 2016, pp. 265--283.

\bibitem{bengio2015deep}
Y.~Bengio, I.~J. Goodfellow, and A.~Courville, ``Deep learning,''
  \emph{Nature}, vol. 521, pp. 436--444, 2015.

\bibitem{salakhutdinov2009deep}
R.~Salakhutdinov and G.~Hinton, ``Deep boltzmann machines,'' in
  \emph{Artificial Intelligence and Statistics}, 2009, pp. 448--455.

\bibitem{roberts2013gaussian}
S.~Roberts, M.~Osborne, M.~Ebden, S.~Reece, N.~Gibson, and S.~Aigrain,
  ``Gaussian processes for time-series modelling,'' \emph{Phil. Trans. R. Soc.
  A}, vol. 371, no. 1984, p. 20110550, 2013.

\bibitem{richardson2017gaussian}
R.~R. Richardson, M.~A. Osborne, and D.~A. Howey, ``Gaussian process regression
  for forecasting battery state of health,'' \emph{Journal of Power Sources},
  vol. 357, pp. 209--219, 2017.

\bibitem{rogers2011adaptive}
A.~Rogers, S.~Maleki, S.~Ghosh, and N.~R. Jennings, ``Adaptive home heating
  control through gaussian process prediction and mathematical programming,''
  2011.

\end{thebibliography}
\newpage

\section*{Appendix A. DMGP Posteriors}
Consider $l^{th}$ hidden matrix-variate GP layer. To lighten the notation, we drop the superscript $l$ during the derivation. Let MGP input $\mathbf{X}$, and MGP output $\mathbf{F}$ \footnote{$\mathbf{X}$ is equivalent to $\mathbf{H}^{l-1}$, while $\mathbf{F}$ is equivalent to $\mathbf{F}^{l}$.}. The joint distribution of $\mathbf{F} \in \mathcal{R}^{N \times P}$ and $\mathbf{U} \in \mathcal{R}^{M \times P}$ is given by:
\begin{equation}
E = p(\mathbf{F}, \mathbf{U}) = \frac{1}{\mathcal{Z}} \exp\Bigg(-\frac{1}{2} \mathrm{tr}\bigg( \mathbf{\Sigma}^{-1} \tilde{\mathbf{M}}^{T} \tilde{\mathbf{K}}^{-1} \tilde{\mathbf{M}} \bigg)\Bigg)
\end{equation}
where $\mathcal{Z} = (2\pi)^{(N+M)P/2} |\mathbf{\Sigma}|^{(N+M)/2} |\tilde{\mathbf{K}}|^{P/2}$, $\tilde{\mathbf{M}} = \Big[\begin{array}{c} \mathbf{F}-m(\mathbf{X}) \\ \mathbf{U}-m(\mathbf{Z}) \\ \end{array}\Big]$, and $\tilde{\mathbf{K}} = \Big[\begin{array}{cc} \mathbf{K}_{XX} & \mathbf{K}_{XZ} \\ \mathbf{K}_{ZX} & \mathbf{K}_{ZZ} \\ \end{array}\Big]$.

Using the equality:
\begin{align}
\tilde{\mathbf{K}}^{-1} &= \Big[\begin{array}{cc} \mathbf{I} & \mathbf{0} \\ -\mathbf{K}_{ZZ}^{-1}\mathbf{K}_{ZX} & \mathbf{I} \\ \end{array}\Big]
\Big[\begin{array}{cc} \mathbf{R}^{-1} & \mathbf{0} \\ \mathbf{0} & \mathbf{K}_{ZZ}^{-1} \\ \end{array}\Big] \nonumber \\
& ~~~~~~ \Big[\begin{array}{cc} \mathbf{I} & -\mathbf{K}_{XZ}\mathbf{K}_{ZZ}^{-1}  \\ \mathbf{0} & \mathbf{I} \\ \end{array}\Big]
\end{align}
where $\mathbf{R} = \tilde{\mathbf{K}}/\mathbf{K}_{ZZ} = \mathbf{K}_{XX} - \mathbf{K}_{XZ}\mathbf{K}_{ZZ}^{-1}\mathbf{K}_{ZX}$ is the Schur complement of $\mathbf{K}$ w.r.t. $\mathbf{K}_{ZZ}$, we can derive that:
\begin{align}
E &= \frac{1}{\mathcal{Z}_1} \exp\Bigg(-\frac{1}{2} \mathrm{tr}\bigg( \mathbf{\Sigma}^{-1} (\mathbf{F}-\mathbf{C})^{T} \mathbf{R}^{-1} (\mathbf{F}-\mathbf{C}) \bigg)\Bigg) \nonumber \\ 
& \times \frac{1}{\mathcal{Z}_2} \exp\Bigg(-\frac{1}{2} \mathrm{tr}\bigg( \mathbf{\Sigma}^{-1} (\mathbf{U}-m(\mathbf{Z}))^{T} \mathbf{K}_{ZZ}^{-1} (\mathbf{U}-m(\mathbf{Z})) \bigg)\Bigg) \label{joint}
\end{align}
where $\mathbf{C} = m(\mathbf{X}) + \mathbf{K}_{XZ}\mathbf{K}_{ZZ}^{-1}(\mathbf{U}-m(\mathbf{Z}))$, $\mathcal{Z}_1 = (2\pi)^{NP/2} |\mathbf{\Sigma}|^{N/2} |\mathbf{R}|^{P/2}$ and $\mathcal{Z}_2 = (2\pi)^{MP/2} |\mathbf{\Sigma}|^{M/2} |\mathbf{K}_{ZZ}|^{P/2}$ given that  $|\tilde{\mathbf{K}}| = |\mathbf{R}||\mathbf{K}_{ZZ}|$.

Note that $E = p(\mathbf{F}, \mathbf{U}) = p(\mathbf{F}|\mathbf{U})p(\mathbf{U})$, and the second term in \eqref{joint} is equivalent to $p(\mathbf{U})$. Thus, conditional distribution of $\mathbf{F}$ given $\mathbf{U}$ is:
\begin{equation}
p(\mathbf{F}|\mathbf{U}) = \mathcal{MN}(\mathbf{F};\mathbf{C}, \mathbf{R},\mathbf{\Sigma}).
\end{equation}

To design scalable Gaussian processes, fully independent training conditional (FITC) approximation \cite{snelson2006sparse, bui2016deep, salimbeni2017doubly} can be applied:
\begin{equation}
p(\mathbf{F}|\mathbf{U}) = \prod_{n=1}^{N} p(\mathbf{f}_n|\mathbf{U}) = \prod_{n=1}^{N} \mathcal{N}(\mathbf{f}_n;\mathbf{c}_n, r_n\mathbf{\Sigma}) .
\end{equation}
By considering observation noise $\mathbf{w}_n \sim \mathcal{N}(\mathbf{0},\mathbf{W})$, the conditional probability distribution of $\mathbf{h}_n = \mathbf{f}_n + \mathbf{w}_n$ is given by:
\begin{equation}
p(\mathbf{H}|\mathbf{F}) = \prod_{n=1}^{N}\mathcal{N}(\mathbf{h}_n; \mathbf{f}_n, \mathbf{W}) .
\end{equation}
\newpage

\section*{Appendix B. Derivation of marginals of the variational posteriors}
Assuming the posterior distribution of each layer is factorized, marginals of the variational posterior is given by:
\begin{align}
q(\{\mathbf{f}^l\}_{l=1}^L) &= \int \prod_{l=1}^{L}p(\mathbf{f}^l|\mathbf{U}^l ;\mathbf{f}^{l-1})q(\mathbf{U}^l) d\{\mathbf{U}^l\}_{l=1}^L \nonumber\\
&= \prod_{l=1}^{L} \int p(\mathbf{f}^l | \mathbf{U}^l;\mathbf{f}^{l-1})q(\mathbf{U}^l) d\mathbf{U}^l \nonumber\\
&= \prod_{l=1}^{L} \mathbb{E}_{q(\mathbf{U}^l)} \left[ p(\mathbf{f}^l | \mathbf{U}^l;\mathbf{f}^{l-1}) \right]
\end{align}

Note from equation \eqref{DMGPform}, the expectation term is simply a linear transformation of a Gaussian, thus resultant distribution also follows a Gaussian $\mathbb{E}_{q(\mathbf{U}^l)} \left[ p(\mathbf{f}^l | \mathbf{U}^l;\mathbf{f}^{l-1}) \right] \sim \mathcal{N}(\mathbf{f}^l;\tilde{\mathbf{c}}^l,\tilde{\mathbf{V}}^l)$ where the mean $\tilde{\mathbf{c}}^l$ is given by:
\begin{equation}
\mathbb{E} \left[ \mathbf{f}^l \right] = m^l(\mathbf{f}^{l-1}) + \boldsymbol{\alpha}^T(\mathbf{A}^l - m^l(\mathbf{z}^l))
\end{equation}
and the covariance $\tilde{\mathbf{V}}^l$ is given by:
\begin{align}
\mathrm{cov} \left[ \mathbf{f}^l, {\mathbf{f}^l} \right] &= (r_n^l \mathbf{\Sigma}^l + \mathbf{W}^l)  +\mathrm{cov} \left[ \mathbf{ \boldsymbol{\alpha}^T U}^l,  \boldsymbol{\alpha}^T  {\mathbf{U}^l} \right] \nonumber \\
&= (r_n^l \mathbf{\Sigma}^l + \mathbf{W}^l) +  \mathbf{\Sigma}^l \mathrm{tr}(\boldsymbol{\alpha}^T S^l \boldsymbol{\alpha}) \nonumber \\
&= (r_n^l + \boldsymbol{\alpha}^T S^l \boldsymbol{\alpha}) \mathbf{\Sigma}^l + \mathbf{W}^l \nonumber \\
&= (k_{f^{l\!-\!1}f^{l\!-\!1}} - \boldsymbol{\alpha}^T(\mathbf{K}_{Z^lZ^l}-\mathbf{S}^l)\boldsymbol{\alpha}) \mathbf{\Sigma}^l + \mathbf{W}^l 
\end{align}
where $ \boldsymbol{\alpha} = \mathbf{k}_{Z^lZ^l}^{-1}\mathbf{K}_{Z^lf^{l\!-\!1}}$. For the last layer, $\mathbf{W}^L$ disappears since $\bar{\mathbf{f}}^L$ actually stands for $\mathbf{f}^L$ instead of $\mathbf{h}^L$ unlike intermediate layers.

\begin{table}[b!]
	\centering
	\caption{Comparison between the ELBO and the MCO}
	\label{ELBO_MCO}
	\begin{tabular}{|c|cc|}
		\hline
		\multicolumn{1}{|c|}{\multirow{2}{*}{Active Zone}} & \multicolumn{2}{c|}{The Lower Bound} \\
		\multicolumn{1}{|c|}{} & DMGPE(ELBO)   & DMGPE(MCO)  \\ \hline
		1  & 13.6   & \textbf{1787.0} \\
		3  & 116.8  & \textbf{1643.9} \\
		4  & 516.7  & \textbf{1957.1} \\
		6  & 409.3  & \textbf{2595.0} \\
		7  & 277.3  & \textbf{1843.2} \\
		9  & 704.0  & \textbf{1800.4} \\
		10 & 184.2  & \textbf{1752.7} \\
		12 & 695.4  & \textbf{1760.8} \\
		13 & 6369.5 & \textbf{8406.4} \\
		15 & 558.0  & \textbf{2142.7} \\
		17 & 2595.5 & \textbf{5933.9} \\
		18 & -218.1 & \textbf{1336.9} \\
		21 & 177.2  & \textbf{2259.0} \\
		23 & -44.8  & \textbf{2412.0} \\
		24 & 1735.2 & \textbf{2693.1} \\
		26 & 3375.7 & \textbf{6179.9} \\
		27 & 96.5   & \textbf{1363.2} \\
		29 & 56.4   & \textbf{1472.2} \\ \hline
	\end{tabular}
\end{table}
\newpage

\section*{Appendix C. Derivation of the MCO}
The model likelihood ($\mathcal{P} \equiv p(\mathbf{Y})$) can be estimated by Monte Carlo estimator \cite{mnih2016variational}:
\begin{equation}
\mathcal{P} = \mathbb{E}_{q(\{\mathbf{U}^l, \bar{\mathbf{F}}_k^l \}_{l=1}^L)} \left[ \Big\langle \frac{p(\mathbf{Y},\{\mathbf{U}^l, \bar{\mathbf{F}}_k^l\}_{l=1}^L)}{q(\{\mathbf{U}^l, \bar{\mathbf{F}}_k^l\}_{l=1}^L)} \Big\rangle \right]
\end{equation}
where $\langle \cdot \rangle$ represents $\frac{1}{K}\sum_{k=1}^K(\cdot)$.

By using \eqref{p_dist} and \eqref{var_dist},
\begin{align}
&\mathcal{P} = \mathbb{E}_{q(\{\mathbf{U}^l, \bar{\mathbf{F}}_k^l \}_{l=1}^L)} \left[ \Big\langle \frac{p(\mathbf{Y},\{\mathbf{U}^l, \bar{\mathbf{F}}_k^l\}_{l=1}^L)}{q(\{\mathbf{U}^l, \bar{\mathbf{F}}_k^l\}_{l=1}^L)} \Big\rangle \right] \nonumber \\
&= \mathbb{E}_{q(\{\mathbf{U}^l, \bar{\mathbf{F}}_k^l \}_{l=1}^L)} \left[ \Big\langle p(\mathbf{Y}|\mathbf{F}_k^L)  \prod_{l=1}^L \frac{p(\mathbf{U}^l)}{q(\mathbf{U}^l)} \Big\rangle \right]  \nonumber \\
&= \mathbb{E}_{q(\{\mathbf{U}^l, \bar{\mathbf{F}}_k^l \}_{l=1}^L)} \left[ \prod_{l=1}^L \frac{p(\mathbf{U}^l)}{q(\mathbf{U}^l)} \Big\langle p(\mathbf{Y}|\mathbf{F}_k^L) \Big\rangle \right] \nonumber \\
&= \mathbb{E}_{q(\{\mathbf{U}^l\}_{l=1}^L)} \left[ \prod_{l=1}^L \frac{p(\mathbf{U}^l)}{q(\mathbf{U}^l)} \mathbb{E}_{q(\{\bar{\mathbf{F}}_k^l\}_{l=1}^L)} \left[ \Big\langle p(\mathbf{Y}|\mathbf{F}_k^L) \Big\rangle \right] \right]
\end{align}
By using Jensen's Inequality, the log likelihood satisfies:
\begin{align}
& \log \mathcal{P} \nonumber \\
& \ge \mathbb{E}_{q(\{\mathbf{U}^l\}_{l=1}^L)} \left[\log \prod_{l=1}^L \frac{p(\mathbf{U}^l)}{q(\mathbf{U}^l)} \mathbb{E}_{q(\{\bar{\mathbf{F}}_k^l\}_{l=1}^L)} \left[ \Big\langle p(\mathbf{Y}|\mathbf{F}_k^L) \Big\rangle \right] \right] \nonumber \\
& = \mathbb{E}_{q(\{\mathbf{U}^l\}_{l=1}^L)} \left[ \log \prod_{l=1}^L \frac{p(\mathbf{U}^l)}{q(\mathbf{U}^l)} \right] \nonumber \\
& ~~~~ + \mathbb{E}_{q(\{\mathbf{U}^l\}_{l=1}^L)} \left[ \log \mathbb{E}_{q(\{\bar{\mathbf{F}}_k^l\}_{l=1}^L)} \left[ \Big\langle p(\mathbf{Y}|\mathbf{F}_k^L) \Big\rangle \right] \right] \nonumber \\
& \ge \mathbb{E}_{q(\{\mathbf{U}^l\}_{l=1}^L)} \left[ \log \prod_{l=1}^L \frac{p(\mathbf{U}^l)}{q(\mathbf{U}^l)} \right] \nonumber \\
& ~~~~ + \mathbb{E}_{q(\{\mathbf{U}^l\}_{l=1}^L)} \left[ \mathbb{E}_{q(\{\bar{\mathbf{F}}_k^l\}_{l=1}^L)} \left[ \log \Big\langle p(\mathbf{Y}|\mathbf{F}_k^L) \Big\rangle \right] \right] \nonumber \\
& = -\sum_{l=1}^L KL(q(\mathbf{U}^l)||p(\mathbf{U}^l)) + \mathbb{E}_{q(\mathbf{F}_k^L)} \left[ \log \Big\langle p(\mathbf{Y}|\mathbf{F}_k^L) \Big\rangle \right] \nonumber \\
& \equiv \mathcal{L}^K(\mathbf{Y}).
\end{align}
Hence the defined MCO, $\mathcal{L}^K(\mathbf{Y})$, is the lower bound of the model log-likelihood. Note further that, the MCO reduces to ELBO when $K=1$.

Since $\mathbb{E}_{q(\mathbf{F}_k^L)} \left[ \log \Big\langle p(\mathbf{Y}|\mathbf{F}_k^L) \Big\rangle \right]$ is equivalent to the IWAE bound, it gets monotonically tighter as $K$ increases \cite{burda2015importance}. Thus, the MCO gives a tighter bound than the traditional ELBO, as demonstrated by empirical study summarized in Table 5.
\begin{equation}
\mathcal{L}(\mathbf{Y})=\mathcal{L}^1(\mathbf{Y}) \le \mathcal{L}^K(\mathbf{Y}) \le \log \mathcal{P}.
\end{equation}
\newpage

\section*{Appendix D. Deep Boltzmann Machine}
DBM is a deep multi-layer network constructed by stacking several restricted Boltzmann machine (RBM) models.
In this section, we briefly summarize \cite{salakhutdinov2009deep}.
RBM is a network of stochastic binary units $\{0, 1\}$, where each RBM has visible units  $v$ and hidden units $h$. In the following, two deep networks are explored to summarize the formulation and training method. Probability distribution of the visible units of DBM is given by 
\begin{equation}
p(\mathbf{v}; \theta) = \sum_{\mathbf{h}} \frac{\exp\left(-E(\mathbf{v}, \mathbf{h}; \theta) \right)}{Z(\theta)},
\label{Eq16}
\end{equation}
where $E(\mathbf{v}, \mathbf{h}; \theta)$ is known as the energy of the system that is defined as 
\begin{equation}
E(\mathbf{v}, \mathbf{h}; \mathbf\theta) = -\mathbf{v}^T \mathbf{W}^1 \mathbf{h}^1 - \mathbf{h}^{1 T} \mathbf{W}^{2} \mathbf{h}^{2}
\label{DBM}
\end{equation}
where $\theta = \{\mathbf{W}^1, \mathbf{W}^2\}$ is a model parameter for 1$^{st}$ and 2$^{nd}$ layer RBM.
DBM can be optimized by approximate maximum likelihood learning. 
By using Jensen's inequality, the lower bound of log-likelihood is given by
\begin{align}
\log (p(\mathbf{v}; \theta)) &\geq \sum_{\mathbf{h}}  q(\mathbf{h} | \mathbf{v}; \mu) \log(p(\mathbf{v}, \mathbf{h}; \theta)) \nonumber \\
~ & ~~~~ - \sum_{\mathbf{h}}  q(\mathbf{h} | \mathbf{v}; \mu) \log( q(\mathbf{h} | \mathbf{v}; \mu)) \nonumber \\
~ & = \mathcal{L}_1 - \mathcal{L}_2.
\label{mf_lowerBound}
\end{align}
where $\mu$ is variational parameters to approximate the true posterior by mean-field approximation:
\begin{align}
q(\mathbf{h} | \mathbf{v}; \mu) & = \prod^{L_1}_{m=1} \prod^{L_2}_{n=1} q(h_{1,m}) q(h_{2,n}) 
\end{align}
with 
\begin{align}
q(h_{1,m}=1) & = \mu_{1,m} \nonumber \\
q(h_{2,n}=1) & = \mu_{2,n}.
\end{align}
Then, solution of \eqref{mf_lowerBound} is numerically solved by:
\begin{align}
\mathcal{L}_1 & = \sum_i \sum_j W_{1,i,j} v_i \mu_{1,j} + \sum_j \sum_k  \mu_{1,j} W_{2,j,k} \mu_{2,k} \nonumber \\
~ & ~~~~ - \log(Z(\theta))
\label{mf_l1} 
\end{align}
and
\begin{align}
\mathcal{L}_2 & =  \sum_{j} \mu_{1,j} \log ( \mu_{1,j} ) + (1-\mu_{1,j}) \log ( 1-\mu_{1,j} ) \nonumber \\  
~ & ~~~~ \sum_k \mu_{2,k} \log ( \mu_{2,k} ) + (1-\mu_{2,k}) \log ( 1-\mu_{2,k} ).
\label{mf_l2} 
\end{align}
For fixed $\theta$, optimal $\mu$ are given by
\begin{align}
\mu_{1,j} & = \text{sigmoid} \left(\sum_i W_{1,i,j} v_i + \sum_k W_{2,j,k} \mu_{2,k} \right) \nonumber \\ 
\mu_{2,k} & = \text{sigmoid} \left(\sum_j W_{2,j,k} \mu_{1,j}\right).
\label{mf_all}
\end{align}

\newpage
\section*{Appendix E. Further Results}
\begin{figure}[t!]
	\includegraphics[width=\columnwidth]{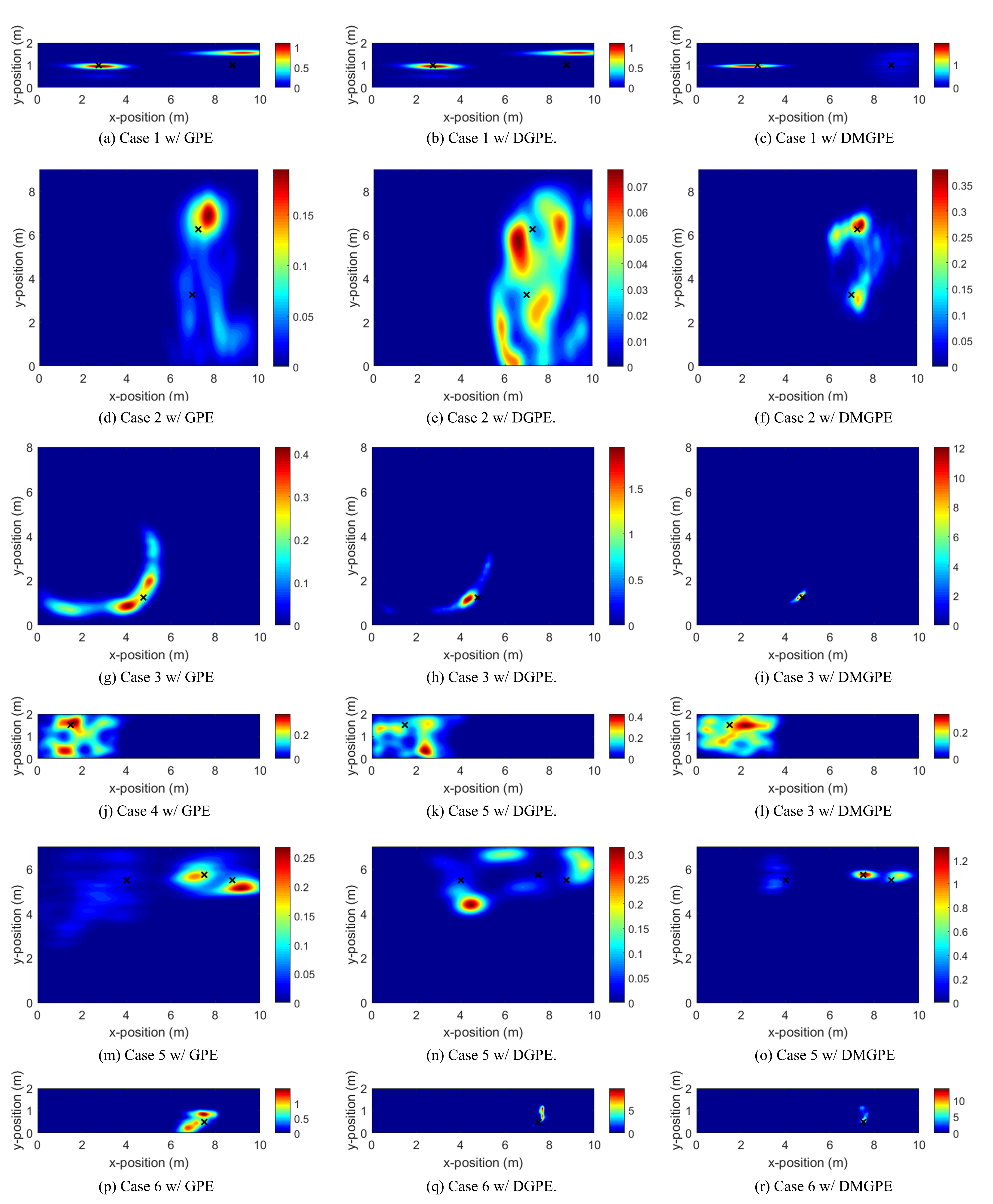}
	\caption{Comparison of contaminant localization results among GP, DGP, and DMGP forecasters at four different design points.}
	\label{fig:loc}
\end{figure}

Fig. \ref{fig:loc} and Table \ref{tab:loc} show contaminant source localization results for the same cases with GPEs, DGPE, and DMGPEs.
DGMPEs outperform to GPEs and DGPEs based source localization results.

\begin{figure}[t!]
	\includegraphics[width=\columnwidth]{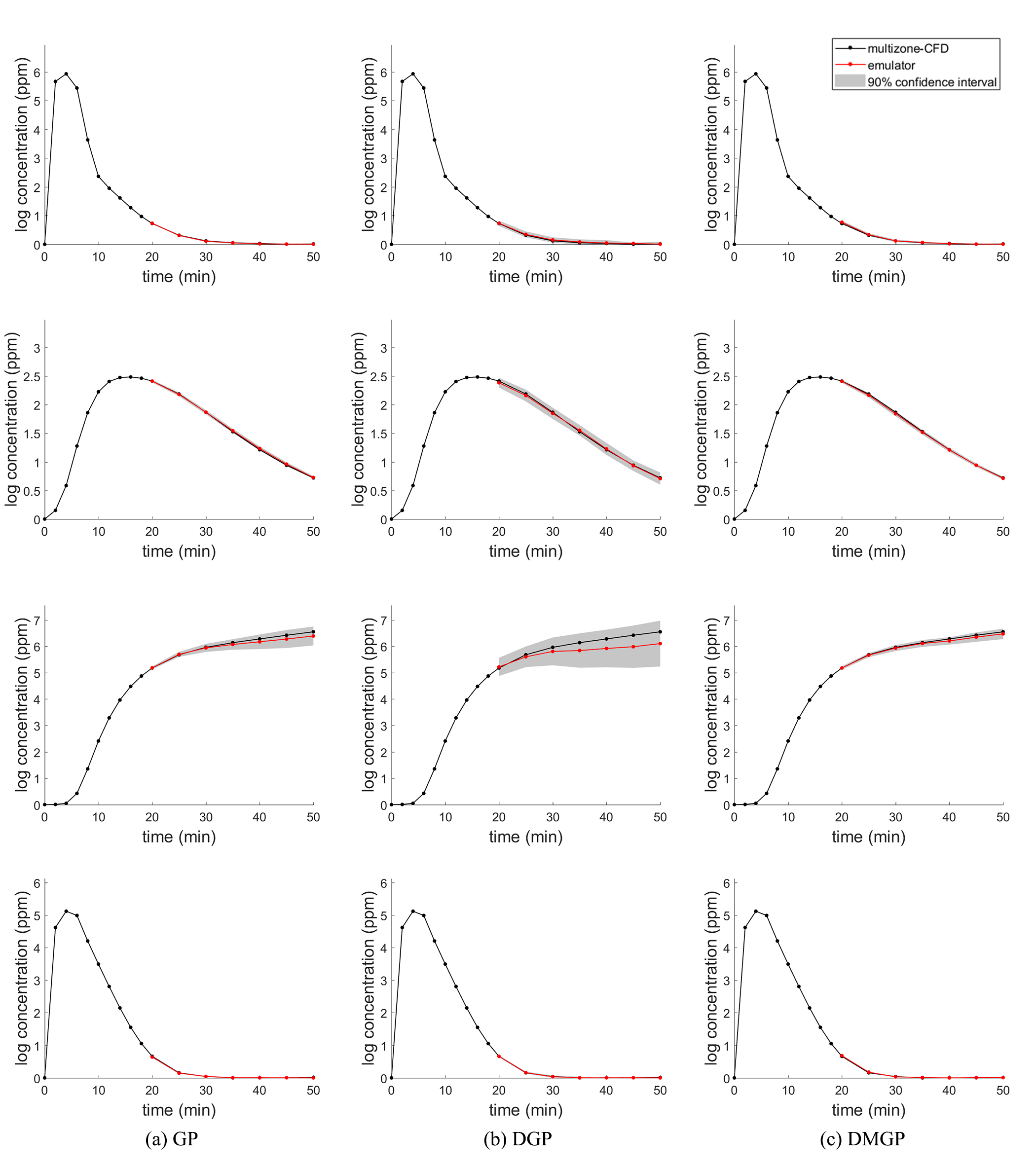}
	\caption{Comparison among GP, DGP, and DMGP forecasters at four different design points.}
	\label{fig:forecast}
\end{figure}

Fig. \ref{fig:forecast}, Table \ref{tab:forecast}, and Table \ref{forecasterMCO}  show forecasting results for the same cases with GPFs, DGPFs, and DMGPFs.
Interestingly, GPFs and DMGPFs do not show significant difference in performance for several tasks, while DGFs often show relatively poor performance.
This is probably due to the low complexity in tasks.\
Unlike emulating tasks, forecasting are relatively easy since the shape of output curves are not highly complex as shown.
Consequentially, too complex structure of DGP disturbed the model training.

\begin{table} [h!]
	\centering
	\caption{Comparison of Contaminant Localization Results Among GPEs, DGPEs, and DMGPEs.}
	\label{tab:loc}
	\begin{tabular}{|c|ccc|ccc|} 
		\hline
		& \multicolumn{3}{c|}{$p(n=n_{true})$} & \multicolumn{3}{c|}{Average location error (m)} \\ 
		~~~ Case ~~~ & GPE & DGPE & DGMPE & GPE & DGPE & DGMPE \\ \hline
		1 & \textbf{1.00}    & \textbf{1.00}    & \textbf{1.00}    & 0.43 & 0.34 & \textbf{0.31} \\
		2 & \textbf{1.00}    & \textbf{1.00}    & \textbf{1.00}    & 0.75 & 0.94 & \textbf{0.56} \\
		3 & \textbf{0.96}    & \textbf{0.96}    & 0.93   	   	   & 1.06 & 0.57 & \textbf{0.13} \\
		4 & 0.88   			 & \textbf{0.93}    & \textbf{0.93}    & 0.56 & 0.58 & \textbf{0.48} \\
		5 & \textbf{1.00}	 & \textbf{1.00}    & \textbf{1.00}    & 0.73 & 0.81 & \textbf{0.31} \\
		6 & 0.63   			 & \textbf{0.96}    & 0.74   		   & 0.37 & 0.41 & \textbf{0.26} \\  \hline
	\end{tabular}
\end{table}

\begin{table} [h!]
	\centering
	\caption{RMSE and Log-likelihood of Forecasting Results OF Test Cases in Fig \ref{fig:forecast}.}
	\label{tab:forecast}
	\begin{tabular}{|c|ccc|ccc|} 
		\hline
		& \multicolumn{3}{c|}{RMSE} & \multicolumn{3}{c|}{log-likelihood} \\ 
		$(z, s)$& GPF & DGPF & DGMPF & GPF & DGPF & DGMPF \\ \hline
		(3,3)   & \textbf{0.024} & 0.070		  & 0.053			 & \textbf{-3877.7} & -3887.5 & -3881.8 \\
		(17,17) & \textbf{0.035} & 0.054		  & \textbf{0.035}   & -3880.6 & -3887.3 & \textbf{-3880.4} \\
		(18,18) & 0.251 	     & 0.800		  & \textbf{0.139}   & -3892.0 & -3901.0 & \textbf{-3889.5} \\
		(29,29) & 0.029 		 & \textbf{0.013} & 0.041 			 & \textbf{-3879.0} & -3879.4 & -3885.6 \\  \hline
	\end{tabular}
\end{table}

\begin{table}[h!]
	\centering
	\caption{The Lower Bound of Trained GPF, DPGF, and DMGPF.}
	\label{forecasterMCO}
	\begin{tabular}{|c|ccc|}
		\hline
		\multicolumn{1}{|c|}{\multirow{2}{*}{Active Zone}} & \multicolumn{3}{c|}{The Lower Bound} \\
		\multicolumn{1}{|c|}{} & GPE    & DGPE   & DMGPE  \\ \hline
		1  & 1588.7          & 440.3   & \textbf{1837.7}  \\
		3  & 1698.4          & 462.1   & \textbf{1847.3}  \\
		4  & 2367.2          & 803.4   & \textbf{2770.8}  \\
		6  & 1904.5          & 759.7   & \textbf{1982.8}  \\
		7  & 2331.3          & 777.1   & \textbf{2639.5}  \\
		9  & 2437.4          & 768.1   & \textbf{2650.4}  \\
		10 & 2367.9          & 733.5   & \textbf{2582.9}  \\
		12 & 2383.3          & 807.2   & \textbf{2524.7}  \\
		13 & -5608.7         & -5726.2 & \textbf{-5578.7} \\
		15 & 2212.6          & 817.6   & \textbf{2361.4}  \\
		17 & 5997.8          & 2855.4  & \textbf{6939.9}  \\
		18 & 856.8           & 195.2   & \textbf{1499.7}  \\
		21 & 1416.1          & 609.1   & \textbf{1522.2}  \\
		23 & 1768.0          & 700.9   & \textbf{1836.1}  \\
		24 & -1807.5         & -1976.5 & \textbf{-1719.6} \\
		26 & -3108.5         & -3242.3 & \textbf{-2997.9} \\
		27 & 610.0           & 197.0   & \textbf{620.9}   \\
		29 & \textbf{1028.7} & 341.1   & 1006.5  \\ \hline
	\end{tabular}
\end{table}

\end{document}